\documentclass[sigconf]{acmart}
\usepackage{subfigure}
\usepackage{multirow}
\usepackage{diagbox}
\usepackage{booktabs}
\usepackage{array}
\usepackage{makecell}
\usepackage{amsmath}
\usepackage{enumerate}

\AtBeginDocument{%
  \providecommand\BibTeX{{%
    \normalfont B\kern-0.5em{\scshape i\kern-0.25em b}\kern-0.8em\TeX}}}

\setcopyright{acmcopyright}
\copyrightyear{2020}
\acmYear{2020}
\acmDOI{xxx}

\acmConference[arxiv '20]{arxiv '20}{2020}{xxx}
\acmBooktitle{xxx}
\acmPrice{xxx}
\acmISBN{xxx}

\begin{document}

\title{The First Step Towards Modeling Unbreakable Malware}

\author{Tiantian Ji}
\email{jitiantian0728@bupt.edu.cn}
\affiliation{%
  \institution{Key Laboratory of Trustworthy Distributed Computing and Service (BUPT), Ministry of Education, Beijing University of Posts and Telecommunications}
  \city{Beijing}
  \state{China}
  \postcode{100876}
}

\author{Binxing Fang}
\email{fangbx@bupt.edu.cn}
\affiliation{%
  \institution{Key Laboratory of Trustworthy Distributed Computing and Service (BUPT), Ministry of Education, Beijing University of Posts and Telecommunications}
  \city{Beijing}
  \state{China}
  \postcode{100876}
}

\author{Xiang Cui}
\email{cuixiang@gzhu.edu.cn}
\authornotemark[1]
\affiliation{%
  \institution{Cyberspace Institute of Advanced Technology, Guangzhou University}
  \city{Guangzhou}
  \state{China}
  \postcode{510006}
}

\author{Zhongru Wang}
\email{wangzhongru@bupt.edu.cn}
\affiliation{%
  \institution{Chinese Academy of Cyberspace Studies}
  \city{Beijing}
  \state{China}
  \postcode{100010}
}

\author{Jiawen Diao}
\affiliation{%
  \institution{Key Laboratory of Trustworthy Distributed Computing and Service (BUPT), Ministry of Education, Beijing University of Posts and Telecommunications}
  \city{Beijing}
  \state{China}
  \postcode{100876}
}

\author{Tian Wang}
\affiliation{%
  \institution{Key Laboratory of Trustworthy Distributed Computing and Service (BUPT), Ministry of Education, Beijing University of Posts and Telecommunications}
  \city{Beijing}
  \state{China}
  \postcode{100876}
}

\author{WeiQiang Yu}
\affiliation{%
  \institution{(Beijing DigApis Technology Co., Ltd}
  \city{Beijing}
  \state{China}
  \postcode{100081}
}


\begin{abstract}
  Constructing stealthy malware has gained increasing popularity among cyber attackers to conceal their malicious intent. 
  Nevertheless, the constructed stealthy malware still fails to survive the reverse engineering by security experts. 
  Therefore, this paper modeled a type of malware with an ``unbreakable'' security attribute-unbreakable malware (UBM), and made a systematical probe into this new type of threat through modeling, method analysis, experiments, evaluation and anti-defense capacity tests. 
  Specifically, we first formalized the definition of UBM and analyzed its security attributes, put forward two core features that are essential for realizing the ``unbreakable'' security attribute, and their relevant tetrad for evaluation. 
  Then, we worked out and implemented four algorithms for constructing UBM, and verified the ``unbreakable'' security attribute based on our evaluation of the abovementioned two core features. 
  After that, the four verified algorithms were employed to construct UBM instances, and by analyzing their volume increment and anti-defense capacity, we confirmed real-world applicability of UBM. 
  Finally, to address the new threats incurred by UBM to the cyberspace, this paper explored some possible defense measures, with a view to establishing defense systems against UBM attacks. 
\end{abstract}

  

\keywords{Unbreakable Malware, Accurate identification, Intent concealment}


\maketitle

\section{Introduction}
Malware comes in various forms, including viruses, worms, remote access Trojans , bots, ransomware, etc. 
Since the Morris worm appeared in 1988, malware has demonstrated its destructiveness and started to cause international concern. 
In recent years, cybersecurity companies and media have released many reports on major cybersecurity incidents, most of which actually focus on malware analysis. 
Undoubtedly, malware accounts for a thorny problem in cyberspace security \cite{tian01}.

As the lasting arms race featuring cyberattack and defense is entering a new stage, many malware programs attempt to conceal their existence and malicious intent, and thus are called ``stealthy malware'' (SM for short). 
This type of malware stands a chance ten times higher than conventional malware to launch a successful attack \cite{DBLP:conf/ndss/WangHLJYZRCCGC20}.
A recent report \cite{David03} estimates that SM-based attacks account for 35\% of all attacks at present, and the attacks in the first half of 2019 alone, for example, increased by 364\%. 
It is not hard to see that constructing stealthy malware has become a trend. 

Despite the lots of studies invested into construction and improvement of stealthy malware,  the malware's malicious intent still risks being uncovered. 
For example, the Stuxnet worm, a notorious stealthy malware program, has been constructed only aimed at attacking industrial control systems that are installed with some particular software and hardware. 
Another noted example of stealthy malware is the DarkHotel spyware which incorporates a function to detect virtual machines, sandbox environment and antivirus engines. These environments are not the target environments of DarkHotel, and only in the case where they are not detected will the spyware perform its malicious acts. 
Analysis of the Stuxnet worm and DarkHotel spyware, however, reveals that their attack targets, malicious intent and attack methods are all hardcoded in malware files. 
The security defenders can employ static analysis to conduct complete reverse engineering of such malware, which is the major challenge that stealthy malware engineers face at present.

DeepLocker \cite{Dhilung04}, a malware that IBM Research proposed on Black Hat 2018, was the first proof-of-concept of the new type of malware that this paper is devoted to. 
Before an attack is unleashed, the DeepLocker realizes the perfect ``unbreakable'' attribute with the help of AI technologies, i.e., even if its codes are available as open source, its malicious intent remains stealthy. 
Moreover, in an actual attack case, the malware BIOLOAD coded by the APT group FIN7, based on the attribute of ``computer name'' and hashing, also attempted to attain this ``unbreakable'' attribute. 
BIOLOAD once again proves that unbreakable malware can be employed as a novel advanced technique to launch attacks in real-world application.

In this connection, this paper modeled a type of UnBreakable Malware (UBM) and analyzed it in an all-round manner through modeling, method analysis, experiments, evaluation and anti-defense capacity tests.  
The main contributions of this paper are as follows: 
\begin{enumerate}[1)]
\item This paper, for the first time, proposed the concept of UBM and modeled it formally. 
The UBM abandons the ``if this, then that'' target identification algorithm that the existing stealthy malware depends on, displacing it with a new algorithm that is able to precisely perceive a (or a type of) target environment; 
meanwhile, it conceals its malicious intent. 
In this way, a common malware can be equipped with the ``unbreakable'' security attribute, thus are able to defend against reverse engineering completely.  
\item	This paper proposed two core features -- non-enumerability and definiteness, which are essential for realizing the ``unbreakable'' security attribute. 
Further, a tetrad to evaluate the two core features was put forward in this paper through formalized definition and analysis. 
The malware that meets the constraint conditions for this tetrad is able to survive forward analysis and reverse engineering. 
Thereby, the two core features work to ensure the realization of the ``unbreakable'' security attribute in a thorough manner.
\item	Based on the formalized definition of UBM and research on its security attributes, this paper then proposed an architecture for constructing UBM, and found four algorithms to realize the two functions of ``accurate identification'' and ``intent concealment''. 
Also, by implementing these four algorithms and evaluating their security, this paper confirmed that the UBM \textit{Discriminator} constructed based on the four algorithms had incorporated this ``unbreakable'' security attribute, as is in conformity with the standard for UBM construction. 
\item	Based on the four algorithms and the UBM six-tuple, this paper has constructed four actual UBMs. 
Evaluation of the applicability of these programs proved that UBM can realize an unperceivable volume increment and successfully avoid the detection of almost all antivirus engines, thereby verifying the high-threat combat capacity of UBM.
\item To defend against the security threats incurred by UBM, this paper has identified the vulnerabilities of UBM during the two links of its construction: ``target attribute designation'' and ``functioning stages'', and explored some possible defense solutions, with a vision to enhance UBM-oriented security defense. 
\end{enumerate}

\section{Background and Related Work}
This section presents relevant research work. 
In view of some existing problems in constructing stealthy malware, the research background, related work, 
and our UBM modeling views are introduced in this section. 

\subsection{Basic concepts}
To conduct accurate UBM modeling, this paper first defines relevant concepts as follows:
\begin{itemize}
  \item \textbf{D1 \textit{Malicious Payload:}} Independent files or code segments that harbor malicious intent, like the code segments in NotPetya that perform the function of erasing the hard drive. 
  Moreover, in this paper, the plaintext attack payload is termed \textit{plain payload} (denoted as ``\textit{pp}''), while the corresponding encrypted ciphertext attack payload is termed \textit{cipher payload} (denoted as ``\textit{cp}'').
  \item \textbf{D2 \textit{Malicious Intent:}} Intent carried by the malicious payload to pinpoint its attack target, technique, act, purpose, etc. 
  Common malicious intent includes intelligence stealing, data encryption, hard-drive erasure, data shredding and so on. 
  \item \textbf{D3 \textit{Target Attribute:}} The attribute designated by an attacker. 
  In this paper, we use \textit{T} to denote a target attribute sample, and its corresponding non-target attribute sample is denoted as $\overline{T}$; 
  the set of target attribute samples and that of the non-target attribute samples together constitute the input space \textit{X} of the UBM. 
  In this paper, the UBM establishes a ``one-to-one'' or ``one type-to-one'' mapping relation between the target attribute sample and the key. 
  This paper defines the target attribute used for ``one-to-one'' mapping as the target attribute of uniqueness, e.g., a specific file or a specific computer name; 
  and the target attribute used for ``one type-to-one'' mapping is defined as the target attribute of unique-typedness, e.g., the facial images  of Tom Cruise. 
\end{itemize}

\subsection{Triggering conditions for stealthy malware attacks}
Before an attack is unleashed, the malware seeks stealthiness by hiding its malicious intent. 
An attack will be launched if and only if the stealthy malware meets some specific attack triggering conditions.
By the attack triggering condition, this paper divides the attack implementation mechanism of existing stealthy malware into two types: target environment detection-based implementation, and complex logic judgment-based implementation. 
\begin{enumerate}[$\bullet$]
  \item \textbf{Target environment detection}
\end{enumerate}
Existing stealthy malware determines whether to unleash an attack by detecting and judging whether the target environment meets its anticipation. Detectable indicators include the residential country, language, operation system (version) and system performance of the current host computer. 
For instance, only when a specific platform or an operation system (version) is found as expected can the malware for Stuxnet, Duqu, Gauss and Flame execute a malicious attack \cite{DBLP:journals/computer/ChenA11}. 

Furthermore, researchers have proposed various sandbox detection techniques \cite{DBLP:conf/sp/MiramirkhaniANP17,DBLP:conf/woot/BlackthorneBFBY16,DBLP:conf/ccs/KolbitschKK11} to help design the attack triggering conditions, and conducted comprehensive detection and evaluation of various sandbox features to determine whether the current host computer environment is the sandbox environment. 
If it is detected as the sandbox environment, the malware will not work or attempt to make the system break down. Only in a non-sandbox environment can the malware be triggered off to unleash an attack, and this non-sandbox environment is actually the target environment of the stealthy malware. 

\begin{figure}[h]
  \centering
  \includegraphics[scale=0.35,trim=0mm 1mm 0mm 0mm,clip=true]{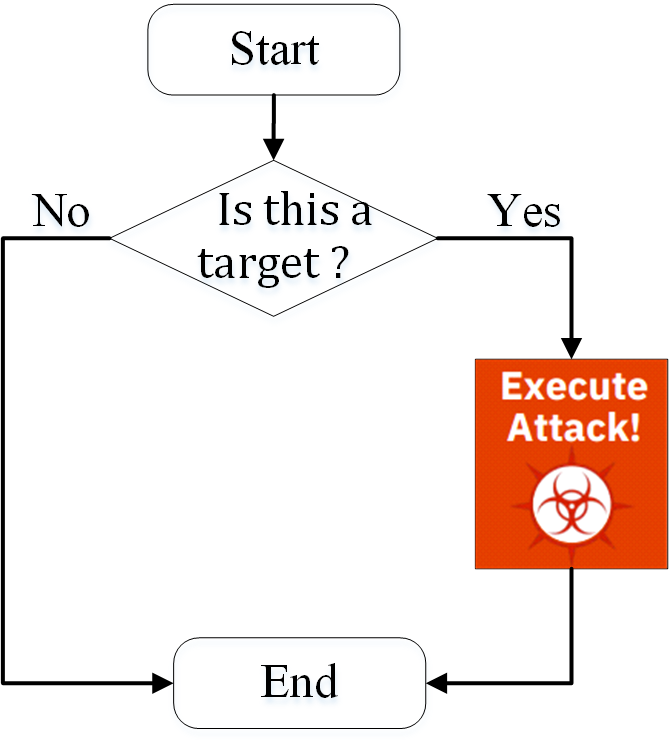}
  \caption{Execution Flow of stealthy malware}
  \Description{The idea is: if this, then that.}
  \label{fig: figure2}
\end{figure}
\begin{figure*}[h]
  \centering
  \includegraphics[scale=0.38,trim=0mm 4mm 0mm 0mm,clip=true]{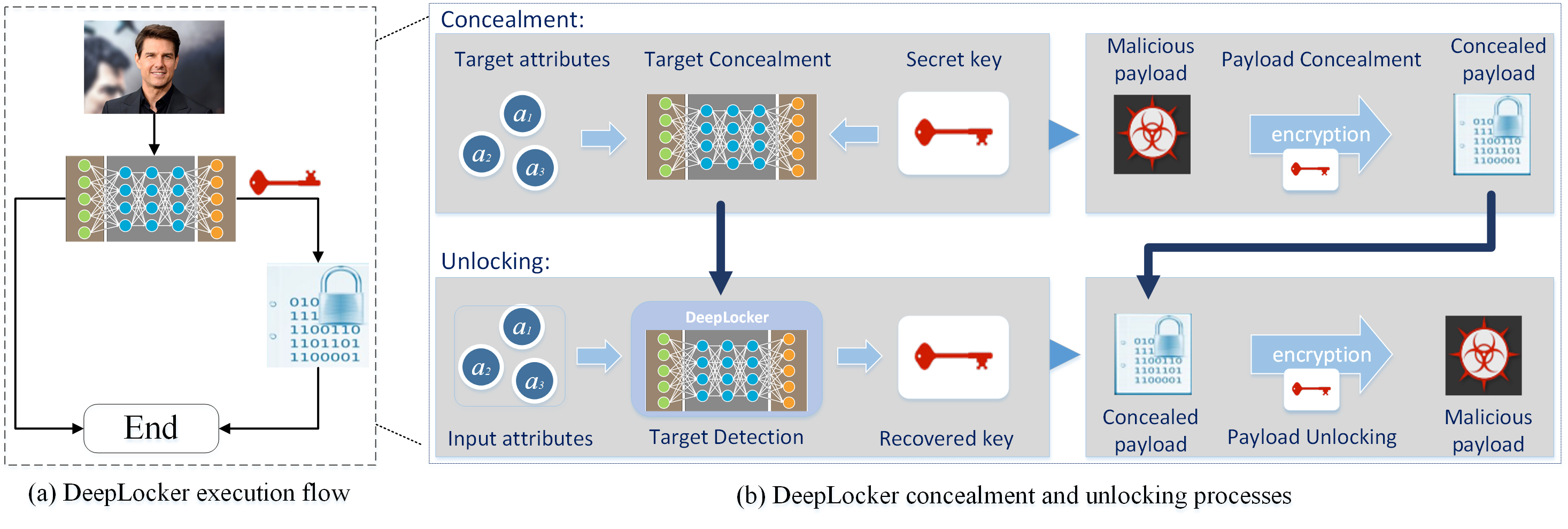}
  \caption{Execution flow of DeepLocker and its concealment and unlocking operations}
  \Description{The left subfigure shows the execution flow of DeepLocker; the right subfigure shows the concealment and unlocking processes of DeepLocker.}
  \label{fig: figure3}
\end{figure*}
\begin{enumerate}[$\bullet$]
  \item \textbf{Complex logic judgment}
\end{enumerate}
One dynamic analysis of malware can only cover one execution path, so complex logic structures are designed for various types of stealthy malware to conceal their execution branches of the malicious payload. Unleashing a malicious attack depends on the execution of the complex logic branches. 
For example, magic bytes or string matching, stalling loop and infinite loop all fall into the category of complex logic judgment \cite{DBLP:books/sp/20/UbaleJ20}. 

%
%
This type of stealthy malware that launches attacks based on complex logic judgment, as presented in the above examples, has made it a challenge in the detection process of symbolic execution \cite{DBLP:conf/ndss/StephensGSDWCSK16}, fuzz testing \cite{afl11} and other dynamic analysis tasks. As a consequence, complex logic judgment-based design is popular among attackers. 

To sum up, in constructing the two attack triggering conditions, we will end up with the judgment ``Is this a target'', no matter how complex the target environment or logic judgment is. 
As shown in Figure~\ref{fig: figure2}, both of the two attack triggering conditions worked out in this paper are in line with the ``if this, then that'' design idea, i.e., as long as the judgment condition ``Is this a target'' is satisfied, the malicious attack will be triggered and unleashed. 

This design, however, has its limitations: 
The information used to crack the attack triggering conditions is always hardcoded in the malware or transmitted via the command and control (C\&C) channel. 
By analyzing the study case of some actual attacks in ATT\&CK \cite{mitre12}, we found that such information can be easily attained through reverse engineering or network monitoring. 
To put it another way, the defender can utilize such information to break the attack triggering conditions and thereby uncover the malicious intent. 

Accordingly, this paper discards the ``if this, then that'' design idea, and replaces it with the two functions of ``accurate identification'' and ``intent concealment''. 
This not only makes it hard to access the information for ascertaining the attack triggering conditions, 
but ensures that the malicious payload will not be decoded even when the code of the malware is available as open source (because the malicious payload encrypted based on this is hard-to-get information). 
In this way, the malicious intent is made ``unbreakable''. 

\subsection{Attack Mechanism of DeepLocker}
In this section, we take DeepLocker as an example of UBM and explore its mechanism to help for modeling UBM for research. 

As shown in Figure~\ref{fig: figure3}, DeepLocker uses a deep neural network (DNN) model to accurately identify the target attributes and conceal the intent of the malicious payload. 
The two core functions are implemented through the two processes -- concealment and unlocking. During concealment, DeepLocker uses the DNN to conduct dynamic concealment of the symmetric key, 
i.e., the key will not be hardcoded into the malware, 
while the input received is taken as a basis for the DNN to determine whether to generate the key or not. 

During concealment, the key is employed to encrypt a plain payload into a cipher payload, 
so the DNN's concealment of the key means successful concealment of the malicious intent. 
A DNN model can help implement our concealment of malicious intent mainly because with its well-trained classification function based on the data sets of target attributes and of non-target attributes, 
it can implement accurate identification of the target attributes.

Correspondingly, during unlocking, the DNN first accurately identifies the target attribute 
(as shown by the facial image of Tom Cruise in Figure~\ref{fig: figure3}) to assist generating the key for decoding the cipher payload, 
and then unleashes an attack. 
In this way, the ``if this, then that'' design idea is displaced.

Based on the attack mechanism of DeepLocker, this paper analyses further the main reasons for its ``unbreakable'' security attribute from the perspective of a security defender: 

First, the target attribute cannot be ascertained by brute-force traversal . The target attribute is assigned as controlled by the attacker, who can make use of specific pictures, files, videos, audios, physical environment, software environment, user action, geographic location, and so on. The defender neither knows the type of the target attribute, nor has any means to match the specific target attribute. It is challenging to conduct brute-force traversal of the target attribute, as hard as or even harder than to crack the AES-128 algorithm, and this is an unreachable hashrate for current electronic computers. Thereby, from the perspective of forward decryption , as the target attribute cannot be ascertained by brute-force traversal, the trigger cannot be activated; consequently, the cipher payload cannot be decrypted and executed, and the malicious intent cannot be identified by the defender. 

Second, the cryptographic algorithm's key strength is at least 128 bits, which is a crucial indicator to measure the security level of a cryptographic algorithm.
Even if DeepLocker is open source, its built-in cipher payload cannot be reversely cracked, mainly in that the DeepLocker takes the AES-128 algorithm by default and has a key strength of 128 bits, and the defender cannot obtain any information by analyzing the open source DeepLocker to crack for the key. Therefore, key guessing  is the only solution for reverse decryption, but no existing techniques are capable of guessing a key with 128 bits strength. 

To sum up, we conclude that the ``unbreakable'' attribute of UBM is determined by two functions -- ``accurate identification'' and ``intent concealment''. On this basis, hereinafter, we perform modeling of the highly stealthy malware and analyze its attributes in the following sections. 

\section{Formalized Modeling of Unbreakable Malware}
\subsection{Assumptions for modeling}
Modeling inevitably depends on specific assumptions. 
To explicitly express the functions and action scope of UBM, this paper made the following assumptions: 

\textbf{Assumption 1:} To demonstrate that the UBM has the ``unbreakable'' security attribute, the UBM in this paper was placed in a cooperative adversarial environment. Advanced antivirus engines were deployed on the perimeter of the network and on the victim host computer for defense, and high-caliber reverse engineers for security were allowed to be present to stay vigilant. 

\textbf{Assumption 2:} As for the application scenario, we assumed that the location of the target was unknown, and the UBM had to seek its attack target by castnet communication. One real-world example of casenet communication is intranet translation  that malware usually employs to seek a target in a physically-isolated private network or organization.  

\textbf{Assumption 3:} During UBM construction, even though the malicious payload supports both symmetric and asymmetric encryption, this paper only assumed symmetric encryption as the algorithm to help construct the UBM, because the asymmetric encryption complicates UBM design and features a low encryption-decryption speed without bringing additional security benefits. Therefore, hereinafter, the ``encryption key'' and ``decryption key'' were jointly called ``key'' and denoted by ``$key$''. Its corresponding wrong key was denoted by ``$\overline{key}$''. 

\textbf{Assumption 4:} From a realistic perspective, this paper assumed that all relevant adversarial attack-and-defense processes related to malware were placed in an electronic computer environment, instead of a quantum computer environment.  

\subsection{Formalized definition of UBM}
Based on the above assumptions, this paper provided a formalized definition of UBM. 

\begin{enumerate}[$\bullet$]
  \item \textbf{Model Define: UBM (\textit{UnBreakable Malware}).} The UBM, comprised of a six-tuple, refers to a type of malware with the ``unbreakable'' security attribute, denoted as \textit{UBM=\{Discriminator, Unb, Judger, Decryptor, cp, BenCode\}}.
\end{enumerate}

\textbf{\textit{Discriminator}:} A \textit{Discriminator} realizes the two core functions of ``accurate identification'' and ``intent concealment'', and converts the input space \textit{X} into \textit{possibleKey}, denoted as \textit{Discriminator}: $X\rightarrow possibleKey$. 
It incorporates the two mapping relations of $T\rightarrow key$ and $\overline{T}\rightarrow \overline{key}$. The function is expressed as $possibleKey=Discriminator(X)$, and then $key=Discriminator(T)$ and $\overline{key}=Discriminator(\overline{T})$. 
For the target attribute of uniqueness, the \textit{Discriminator} represents a tuple of itself, denoted as $Discriminator=\{Discriminator\}$; for the target attribute of unique-typedness, the \textit{Discriminator} represents a two-tuple, denoted as $Discriminator=\{fCollector, Discriminator\}$.

\textbf{\textit{fCollector}:} The feature collector. Aimed at the target attribute of unique-typedness, the \textit{fCollector} extracts the main feature information of the target attribute, helping the \textit{Discriminator} to map a type of target attributes to a unique stable key. 

\textbf{\textit{Unb}:} The intrinsic and ``unbreakable'' security attribute of the UBM. This attribute entails two core features of the malware--``non-enumerability'' and ``definiteness'', and it is denoted as \textit{Unb=\{non-enumerability, definiteness\}}. 

\textbf{\textit{non-enumerability}:} The input space and output space are too big to be traversed in a brute-force manner. All input and output spaces mentioned in this paper refer to the input and output spaces of the \textit{Discriminator}. That is, the input space \textit{X} consists of $T$ and $\overline{T}$, and the output space \textit{possibleKey} comprises $key$ and $\overline{key}$. 

\textbf{\textit{definiteness}:} The algorithm that processes inputs of the correlation engine features definiteness. That is, the same input or the same type of inputs, if fed into the engine for multiple times, will yield the same output.  

These two core features (non-enumerability and definiteness)  are manifested as the functions of ``accurate identification'' and ``intent concealment'' in malware implementation. Thus, the security attribute of UBM depends on the \textit{Discriminator} for realization. 

\textbf{\textit{Judger}:} The security attribute judging function, which works to ensure that the implementation of the \textit{Discriminator} realizes the security attributes of UBM. It takes the \textit{Discriminator} as the input, and outputs the tuple for evaluating the \textit{Discriminator}, denoted as $eva=Judger(Discriminator)$. 

\textbf{\textit{eva}:} A constraint tuple that evaluates the \textit{Discriminator}, expressed as Formula~(\ref{con: equation1}), where \textit{x}, \textit{y}, \textit{z} and \textit{w} represent the critical values to meet the evaluation standards. These values are specified by the attacker as per the actual attack scenario, attack demand and attack capacity. Only when the \textit{eva} evaluation standards are met will the \textit{Discriminator} be considered qualified for constructing the UBM.
\begin{equation}
  eva=
   \left\{P_{in}\leq x, P_{out}\leq y, P_{sta}\geq z, P_{acc}\leq w\right\}
  \label{con: equation1}
\end{equation} 

\textbf{\textit{Decryptor}:} The \textit{Decryptor} takes the cipher payload and \textit{possibleKey} as inputs and attempts to decode the cipher payload. It is denoted as $Decryptor=\{keyJudger, decFunc\}$. 

\textbf{\textit{keyJudger}:} The key judgment mechanism. It judges whether a \textit{possibleKey} is a $key$; when the \textit{possibleKey} is a $\overline{key}$, it will interrupt the execution of the \textit{Decryptor}. 

\textbf{\textit{decFunc}:} A decryption function. If and only if a \textit{possibleKey} is a key can the \textit{decFunc} convert the cipher payload into the plain payload, which is denoted as: $pp=decFunc(cp,key)$; otherwise, the \textit{decFunc} will not produce any output, thereby lowering the risk of UBM's being detected. 

\textbf{\textit{cp}:} The cipher payload. It represents encrypted plain payload, which is obtained through encryption of the key generated by the \textit{Discriminator} based on the target attribute sample \textit{T}. 

\textbf{\textit{BenCode}:} A benign functional code. The UBM, if not detected, inserts the malicious code into the \textit{BenCode} to circumvent detection by an antivirus engine. One common example of \textit{BenCode} is video conferencing software.

\begin{table*}[!htbp]
  \caption{Formalized definitions of non-enumerability and definiteness}
  \label{tab:table1}
  \centering
  \begin{tabular}{ccccc}
   \Xhline{0.75pt}
   \multirow{2}*{\diagbox[innerwidth=2cm]{$T$}{$Unb$}}& \multicolumn{2}{c}{\textsl{\textbf{non-enumerability}}} &\multicolumn{2}{c}{\textsl{\textbf{definiteness}}}\\
   \Xcline{2-5}{0.45pt}
   &\textbf{Input space}&\textbf{Output space}&\textbf{Key stability}&\textbf{Key accessibility}\\
   \Xhline{0.5pt}
   \specialrule{0em}{2pt}{2pt}
   \textbf{uniqueness} & 
    \begin{minipage}{1.0in}
    \begin{equation}
      \begin{aligned}
        P_{in}&=\dfrac{target\ sample}{total\ samples}\\ 
        &=\dfrac{1}{total\ samples} 
      \end{aligned}
      \label{con: equation2}
      \tag{2}
    \end{equation}
    \end{minipage}
    &
    \multirow{5}{*}{
      \begin{minipage}{1.1in}
        \begin{equation}
          \begin{aligned}
            P_{out}&=\dfrac{target\ sample}{total\ samples}\\ 
            &=\dfrac{1}{total\ samples} 
          \end{aligned}
          \label{con: equation4}
          \tag{4}
        \end{equation}
      \end{minipage}
    }
    &
    \multirow{5}*{
     \begin{minipage}{1.8in}
      \begin{equation}
        \begin{aligned}
          P_{sta}=\dfrac{max(m_{S_{0}},m_{S_{1}},m_{S_{2}},\dots,m_{S_{n}})}{\sum_{i}^{n}m_{S_{i}}} 
        \end{aligned}
        \label{con: equation5}
        \tag{5}
      \end{equation}
    \end{minipage}
    }
    &
    \multirow{5}*{
     \begin{minipage}{1.6in}
      \begin{equation}
        \begin{aligned}
          P_{acc}=\dfrac{\sum_{i=0}^{n}((S_{i}\odot key)*m_{S_{i}})}{\sum_{i}^{n}m_{S_{i}}} 
        \end{aligned}
        \label{con: equation6}
        \tag{6}
      \end{equation}
    \end{minipage}
    }
    \\ 
   \specialrule{0em}{2pt}{2pt}
   \cline{1-2}
   \specialrule{0em}{2pt}{2pt}
   \textbf{unique-typeness} &
    \begin{minipage}{1.1in}
      \begin{equation}
        \begin{aligned}
          P_{in}&=\dfrac{target\ sample}{total\ samples}\\ 
          &=\dfrac{O(target\ class)}{O(total\ classes)} 
        \end{aligned}
        \label{con: equation3}
        \tag{3}
      \end{equation}
    \end{minipage}
   \\
   \specialrule{0em}{2pt}{2pt}
   \Xhline{0.75pt}
  \end{tabular}
\end{table*}
\subsection{The two core features}
As specified in the formalized definition of UBM, ``non-enumerability'' and ``definiteness'' are the two core features to satisfy the ``unbreakable'' security attribute (i.e., \textit{Unb}). Table~\ref{tab:table1} presents the formalized analysis of these two features in different scenarios. 
\begin{enumerate}[\textbf{a)}]
  \item \textbf{Non-enumerability} 
\end{enumerate}

This paper gave a formalized definition of non-enumerability in a reversed manner. That is, this paper provided a formalized definition of the possibility to enumerate the target objects, and a smaller possibility of enumeration would mean higher non-enumerability. 
  
The enumerated object(s) in the input space was (were) one or one type of target sample in correspondence with the target attribute. In this paper, the enumerability in the input space is denoted as $P_{in}$. As shown in Formula~(\ref{con: equation2}) and ~(\ref{con: equation3}), $P_{in}$ is the ratio of the number of target samples to that of total samples. 
  
The target attribute of uniqueness has only one corresponding target sample in the input space, so as shown in Formula~(\ref{con: equation2}), the value of $P_{in}$ depends on the number of total samples in the input space. 
  
The target attribute of unique-typedness has only one specific type of target samples in the input space. As shown in Formula~(\ref{con: equation3}), in the input space, we denote the complexity of target samples is denoted as $O(target\ class)$, and the complexity of total samples as $O(total\ classes)$. Then, $P_{in}$ is the ratio of $O(target\ class)$ to $O(total\ classes)$. Suppose each class has the same number of samples, then the value of $P_{in}$ will be dependent on the size of the sample class space in the input space. 
  
In the output space, $P_{out}$ is used to express the possibility of enumerated targets. The target object that we enumerate is a key, which is unique. Thus, as shown in Formula~(\ref{con: equation4}), the value of $P_{out}$ will be dependent on the number of total samples in the output space. What's more, as the key is in the plain text format, we suppose the cipher payload is open source, then the number of total samples in the output space will be dependent on the key strength. 
\begin{enumerate}[\textbf{b)}]
  \item \textbf{Definiteness}
\end{enumerate}

Definiteness helps to ensure the quality of the key generated by the \textit{Discriminator}. In this paper, a high-quality key is termed a ``stable key''.  To generate stable key, this paper divides ``definiteness'' further into two important probability attributes for formalized definition. They are ``key stability'' and ``key accessibility'', respectively denoted as $P_{sta}$ and $P_{acc}$. 

\begin{enumerate}[\textcircled{1}]
  \item \textbf{Key stability}
\end{enumerate}

This paper prescribes the following essential requirements for key stability: most of the positive samples must have the same corresponding key within an acceptable scope of fault tolerance. In this paper, the target attribute input samples are positive samples, while those non-target attribute samples are called negative samples.   

Hereby, we have the formalized definition of $P_{sta}$, as shown in Formula~(\ref{con: equation5}), and $P_{sta}$ comes out as a probability. When $P_{sta}$ approaches 0, the key has extremely low stability; when $P_{sta}$ approaches 1, the key has good stability. 

In Formula~(\ref{con: equation5}), $S_i$ refers to the i-th \textit{possibleKey} mapped with positive samples, and $m_{S_i}$ refers to the count of $S_i$, i.e., the number of positive samples that correspond to the same \textit{possibleKey}. A larger $m_{S_i}$ means higher stability of $S_i$. Thus, we chose the maximum value $m_{S_x}$, and took the corresponding $S_x$ as the key. Thereby, $P_{sta}$ would be the ratio of the number of samples of the corresponding key to the number of all positive samples. Therefore, the bigger $P_{sta}$ is, the more stable the key will be, as is more in line with the requirements in this paper. 

\begin{figure*}[!t]
  \centering
  \subfigure[Concealment of malicious intent] {
    \includegraphics[scale=0.38,trim=0mm 2mm 0mm 0mm]{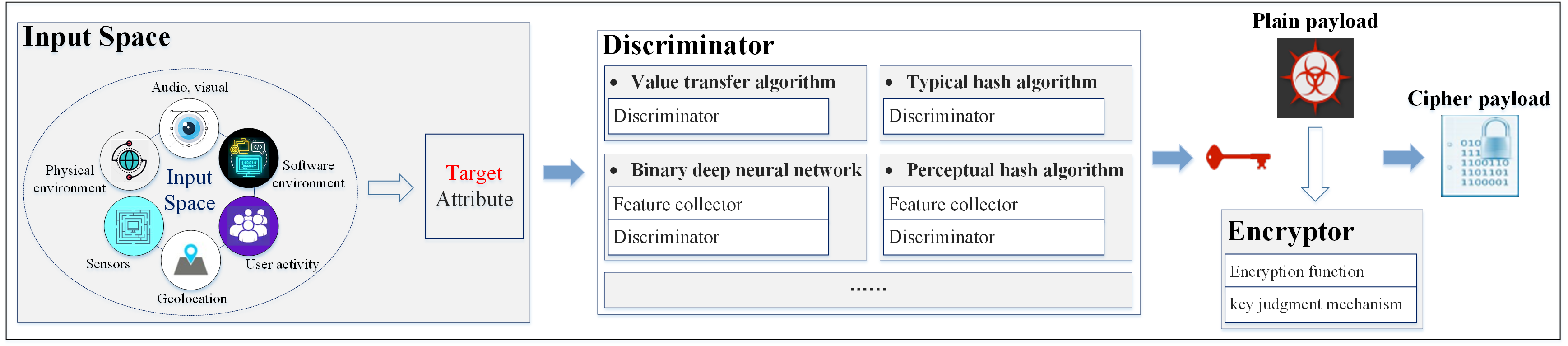}
  }
  \subfigure[Unlocking of malicious intent] {
    \includegraphics[scale=0.38,trim=0mm 2mm 0mm 0mm]{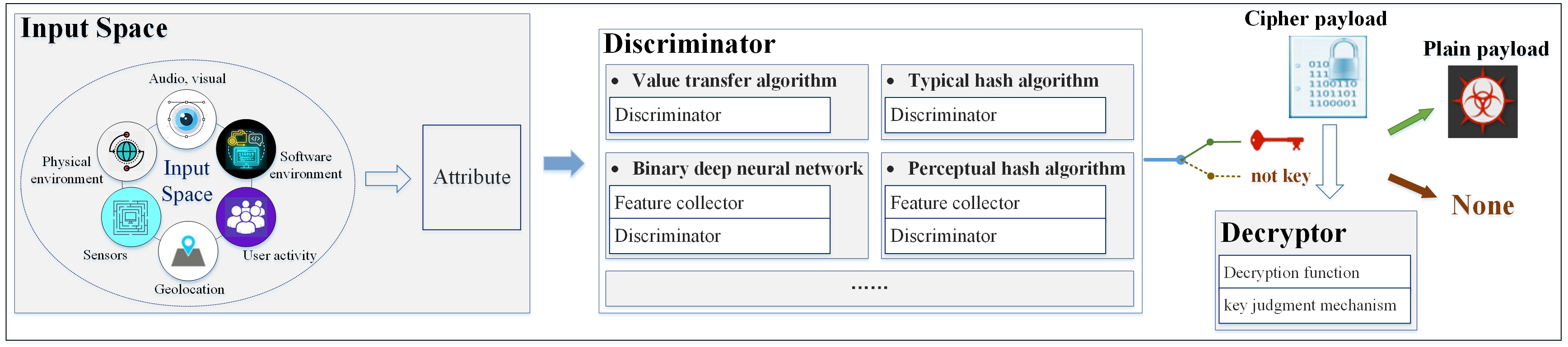}
  }
  \caption{Architecture for UBM construction}
  \Description{The architecture of UBM contains two operation processes that are concealment and unlocking, aiming to build the UBM instances.}
  \label{fig: figure4}
\end{figure*}
\begin{enumerate}[\textcircled{2}]
  \item \textbf{Key accessibility}
\end{enumerate}
This paper had the following essential requirements for key accessibility: Most of the negative samples should not be able to generate a key (i.e., the accessibility of key approaches 0) within an acceptable scope of fault tolerance. 

In this paper, $P_{acc}$ has a formalized definition as shown in Formula~(\ref{con: equation6}). As with $P_{sta}$,  $P_{acc}$ also comes out as a probability. When $P_{acc}$ approaches 0, the negative samples can hardly be mapped to the key; when it approaches 1, almost all negative samples can be mapped to the key through the \textit{Discriminator}. 

In Formula~(\ref{con: equation6}), $S_i$ refers to the i-th output mapped from negative samples. If $S_i$ is the same as the key, the key is accessible, and $S_i$ is an accessible sample. Otherwise, the key is not accessible, and $S_i$ is an inaccessible sample. 
Accordingly, $\sum_{i=0}^{n}((S_{i}\odot key)*m_{S_{i}})$
represents the count of accessible samples. $P_{acc}$ is the ratio of the count of all accessible samples to that of all negative samples. The smaller  $P_{acc}$ is, the less accessible the key will be, as is more in line with the requirements in this paper.

\subsection{UBM construction architecture}
Based on the formalized definition of UBM and our research on its security attributes, we proposed an architecture for UBM construction. It comprises two processes of malicious intent concealment and unlocking, as shown in Figure~\ref{fig: figure4}. 

During the concealment proccess, the attacker assigns the target attribute in the input space \textit{X}, and uses a \textit{Discriminator} in conformity with the UBM construction criteria to generate a key. The encryptor, with the key and the plain payload as its input, outputs the cipher payload and thereby conceals the malicious intent. 

In the unlocking process, the UBM does not know its attack target, so the built-in \textit{Discriminator} takes possible attribute samples as the input, and outputs \textit{possibleKey} to perform decryption over and over again. Because of its accurate identification function, when the \textit{Discriminator} identifies the target attribute, the condition $possibleKey=key$ is satisfied, after which the cipher payload can be decrypted and the plain payload can be generated. 
Meanwhile, because of the intent concealment function of the \textit{Discriminator}, when it has not identified the target attribute, the output $possibleKey=\overline{key}$ , thereby concealing the key. In this case, as there is no way to decrypt the cipher payload, concealment of key is equivalent to concealment of the malicious intent.  

\begin{figure}[h]
  \centering
  \includegraphics[scale=0.35,trim=0mm 1mm 0mm 0mm,clip=true]{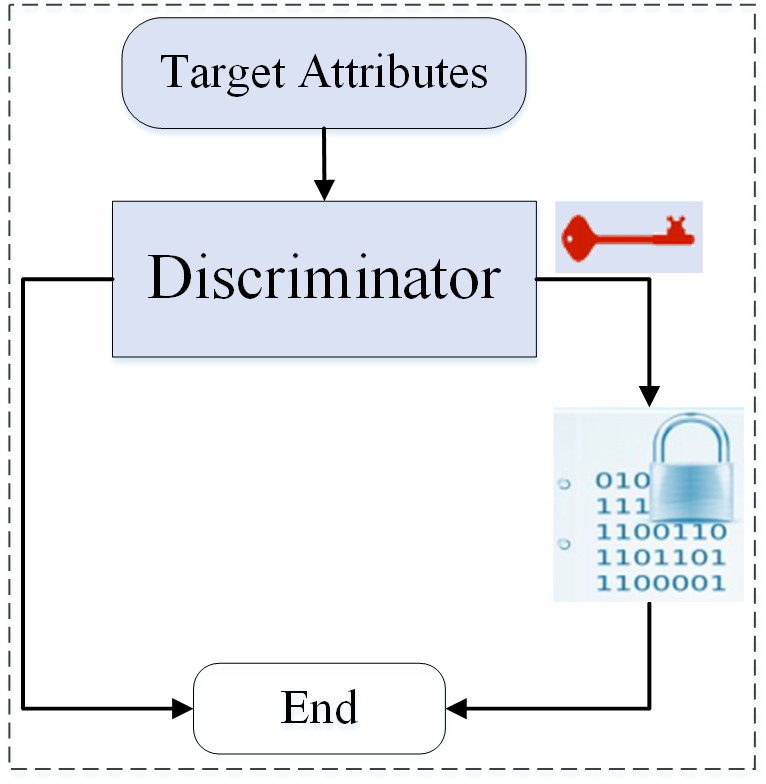}
  \caption{Execution Flow of UBM}
  \Description{Two core functions are: accurate recognition and intent hiding.}
  \label{fig: figure5}
\end{figure}
Figure~\ref{fig: figure5} shows the implementation flow of the UBM when the target attribute is identified. 
In comparison with the SM implementation flow (Figure~\ref{fig: figure2}), UBM replaces the ``if this, then that'' attack-triggering idea with the realization of the \textit{Discriminator}'s functions -- ``accurate identification'' and ``intent concealment''. 
To realize these two functions, the following requirements must be met: 
(1) The input space of the \textit{Discriminator} should be non-enumerable. 
(2) In the mapping relation implemented by the \textit{Discriminator}, the key will be output when and only when the input is the target attribute. 
(3) The input space should also be non-enumerable; in other words, the output key must be at least 128-bit strong. When these three requirements are met, the UBM will be unbreakable and survive both forward analysis or reverse engineering. 
This clearly demonstrates the necessity of ``non-enumberabilty'' and ``definiteness''. 

\begin{table*}[!htbp]
  \caption{Tuples for modeling four UBM}
  \label{tab:table2}
  \begin{tabular}{@{}ccccccc@{}}
  \toprule
  \multicolumn{7}{c}{\textit{\textbf{\begin{tabular}[c]{@{}c@{}}UBM=\{Discriminator, Unb, Judger, Decryptor, cp, BenCode\}\end{tabular}}}}                                                                    \\ \midrule
  Number & \textit{\textbf{Discriminator}} & \textit{\textbf{Unb}} & \textit{\textbf{eva$\leftarrow$Judger}} & \textit{\textbf{Decryptor}} & \textit{\textbf{cp}} & \textit{\textbf{BenCode}} \\ \midrule
  \specialrule{0em}{4pt}{4pt}
  \textcircled{\scriptsize 1} & Value transfer algorithm     & \multirow{12}{*}{Unbreakable}     & 
    \multirow{2}{*}{
      \(
        \left\{
          \begin{aligned}
          P_{in}\leq \dfrac{1}{2^{128}} & \\
          P_{out}\leq \dfrac{1}{2^{128}} & \\
          P_{sta}=1 & \\
          P_{acc}=0 & 
          \end{aligned}
        \right\} 
      \)
    }&
      \(\begin{aligned}
        \varphi &= PKCS7Padding \\
        \delta &= AES-128
       \end{aligned}
      \)& 
    \multirow{12}{*}{\begin{tabular}[c]{@{}p{1.8cm}@{}}In this paper, a.exe in trickbot is taken as \textit{pp}, while \textit{cp} represents the encrypted output of \textit{Discriminator}.\end{tabular}}    & 
    \multirow{12}{*}{\begin{tabular}[c]{@{}p{1.8cm}@{}}In this paper, \textit{BenCode} is realized as a simple pop-out program.\end{tabular}}       \\ \specialrule{0em}{4pt}{4pt} \cmidrule{5-5}
  \specialrule{0em}{4pt}{4pt}
  \textcircled{\scriptsize 2} & Typical hash algorithm       &&& 
    \(
      \begin{aligned}
        \varphi &= PKCS7Padding \\
        \delta &= AES-256
      \end{aligned}
    \)
    &&\\ \specialrule{0em}{4pt}{4pt} \cmidrule{4-5}
  \specialrule{0em}{4pt}{4pt}
  \textcircled{\scriptsize 3} & Binary deep neural network   && 
    \multirow{2}{*}{
      \(
        \left\{
          \begin{aligned}
            P_{in}\leq \dfrac{1}{2^{128}} & \\
            P_{out}\leq \dfrac{1}{2^{128}} & \\
            P_{sta}\geq 95\% & \\
            P_{acc}\leq 0.5\% & 
          \end{aligned}
        \right\}
      \)
    }&
    \(\begin{aligned}
      \varphi &= PKCS7Padding \\
      \delta &= AES-128
     \end{aligned}
    \)    
    &&\\ \specialrule{0em}{4pt}{4pt} \cmidrule{5-5}
  \specialrule{0em}{4pt}{4pt}
  \textcircled{\scriptsize 4} & Perceptual hash algorithm &&&
  \(\begin{aligned}
    \varphi &= PKCS7Padding \\
    \delta &= AES-128
   \end{aligned}
  \)                           
  &&\\ \specialrule{0em}{4pt}{4pt} \bottomrule
  \end{tabular}
\end{table*}
To sum up, the focus of UBM construction is on the \textit{Discriminator}. 
Based on the attribute judgment function, this paper explored and found the four algorithms that could support construction of the \textit{Discriminator} (Figure~\ref{fig: figure4}): \textcircled{1} Value transfer algorithm; \textcircled{2} Typical Hash algorithm; \textcircled{3} Binary deep neural network; and \textcircled{4} Perceptual Hash algorithm. 
In the following experiments in this paper, we would first construct a \textit{Discriminator} with the abovementioned two core functions based on these four algorithms for construction of the UBM.

\section{Experiment}
Table~\ref{tab:table2} shows the tuples of four UBMs. 
Based on the above-mentioned four algorithms (Figure~\ref{fig: figure4}) that were used to construct the \textit{Discriminator}, we constructed the UBMs to conduct further performance evaluation.  

The experiment was performed on the Windows platform, though the construction of UBM can also be fulfilled on Linux or other platforms. Besides, when constructing the UBM, we used the AES symmetric encryption algorithm, in which the secret key length employed equals to the secret key strength. As a 128-bit key is hard to crack, we made sure that the output key should be at least 128 bits long when constructing the \textit{Discriminator} . 

\subsection{Value transfer algorithm}
The value transfer algorithm is the most simple and direct approach to convert the target attribute into a key. It constructs the \textit{Discriminator} by combination, splicing, clipping, value assignment and other operations, and establishes a mapping from the target attribute to the key. 

In this paper, we chose SSID as the target attribute. As SSID can set be set at any length within the (0, 256] bit zone, we used the combined information of SSID and GUID to generate the key, as specified in formula~(\ref{con: equation7}). 

where $key_{wlan}$ refers to the key for encrypting and decrypting the malicious payload, $len(SSID)$ refers to the length of SSID, $key_{len(SSID)}(SSID)$ is part of the key generated by SSID, 
and \\
$key_{128-len(SSID)}(GUID)$ 
is part of the key generated by GUID, and ``+'' denotes ``combination'' or ``splicing''. 

Specifically, Formula~(\ref{con: equation7}) provides two different schemes to generate the key, depending on the length of SSID: 
\begin{equation}
  \resizebox{.9\hsize}{!}{$
  key_{wlan}=
    \left\{
      \begin{aligned}
          key_{len(SSID)}(SSID)+key_{128-len(SSID)}(GUID), (i) & \\
          md5(SSID), (ii) & 
      \end{aligned}
    \right. \tag{7}
  $}
  \label{con: equation7}
\end{equation}

\begin{enumerate}[(i).]
  \item When the SSID length <=128 bits, we would use all SSID binary representations as part of the key, and the remaining part of the 128 bits would be supplemented by the key generated by GUID. In this case, the final key is a combination of GUID and SSID information.
  \item When the SSID length >128 bits, we performed md5 computing on SSID to generate a 128-bit key. 
\end{enumerate}
\begin{figure}[h]
  \centering
  \includegraphics[scale=0.35,trim=0mm 1mm 0mm 0mm,clip=true]{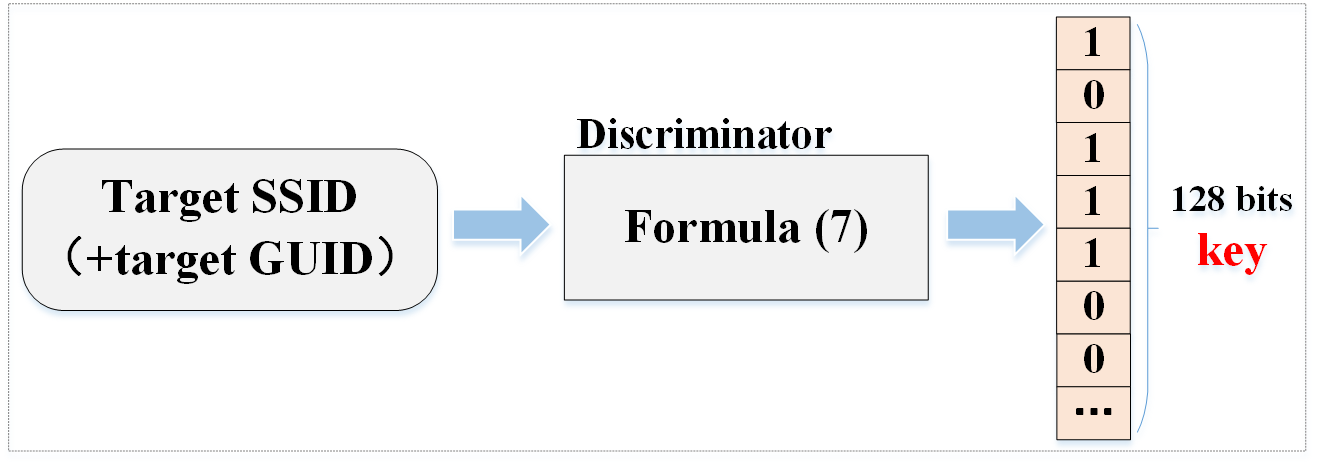}
  \caption{Value transfer algorithm based Discriminator}
  \Description{Value transfer algorithm based Discriminator}
  \label{fig: figure6}
\end{figure}

Finally, we constructed a value transfer \textit{Discriminator} based on Formula~(\ref{con: equation7}). As shown in Figure~\ref{fig: figure6}, when the \textit{Discriminator} detects that the current host computer is located in the target network environment (labeled ``target SSID''), the key will be generated by the value transfer algorithm.

It should be noted that the type of target attributes corresponding to the value transfer \textit{Discriminator} are generally short texts. In addition to SSID, the texts can also be a hard disk serial number or a computer name that marks the physical environment of the host computer. They can also be taken as target attributes to construct \textit{Discriminator}.  

\subsection{Typical hash algorithm}
\begin{figure}[h]
  \centering
  \includegraphics[scale=0.35,trim=0mm 1mm 0mm 0mm,clip=true]{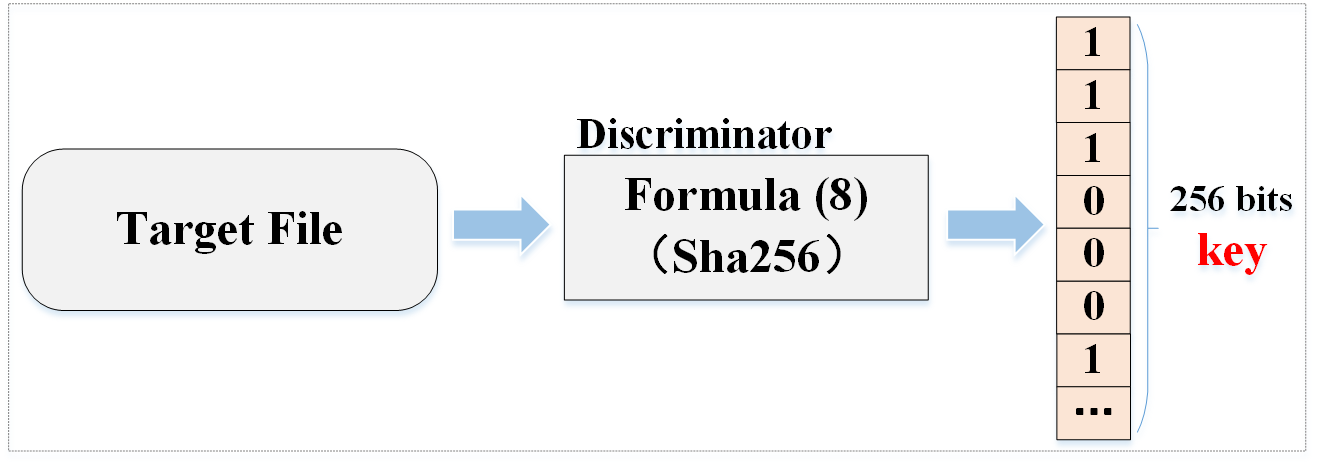}
  \caption{Typical hash algorithm based Discriminator}
  \Description{Typical hash algorithm based Discriminator}
  \label{fig: figure7}
\end{figure}

The key generation scheme (ii) provided by Formula~(\ref{con: equation7}) is designed based on a typical hash algorithm, 
which includes sha1, sha256, sha512, aside from md5. 
These algorithms have some salient features: 
First, the target attribute in these algorithms can be short text objects in the value transfer algorithm, 
or part of the long text (no shorter than 128 bits) included in some specific files in a victim host computer. 
Second, these algorithms can convert the target attributes into unbreakable secret key of different lengths. 
For instance, md5, sha1, sha256 and sha512 algorithms can respectively convert the following secret key lengths: 
128 bits, 160 bits, 256 bits and 512 bits. 
To highlight these two features, this paper took some specific files as the target attributes and constructed a \textit{Discriminator} based on the sha256 algorithm, 
as shown in Formula~(\ref{con: equation8}):
\begin{equation}
  key_{hash}=sha256(tFile) \tag{8}
  \label{con: equation8}
\end{equation}

where \textit{tFile} refers to a designated target file, and \textit{sha256()} can be realized by the sha256() method in the hashlib, and thus $key_{hash}$ is the key generated based on the target file. 
As the example shown in Figure~\ref{fig: figure7}, the APT1 report titled \textit{Appendix C (Digital) - The Malware Arsenal.pdf} is the designated target file; 
if the \textit{Discriminator} takes this file as the input, the output key will be: 0xe22e8ccf50d9e0013688\\229ffbffb4bc3a77e6e46b23726fd83925ba5899af3e.

\subsection{Binary deep neural network}
The binary classification deep neural network (``B-DNN'' for short) model is implemented, aimed at the target attribute of the unique-typedness. During its implementation, the input space should be provided with a type of high-quality target attribute samples set and a type of high-quality non-target attribute samples set to meet the demand for B-DNN training. 
A well-trained B-DNN can establish a ``multiple-to-one'' mapping relationship between one type of target samples and the stable key.

\begin{figure}[h]
  \centering
  \includegraphics[scale=0.35,trim=0mm 1mm 0mm 40mm,clip=true]{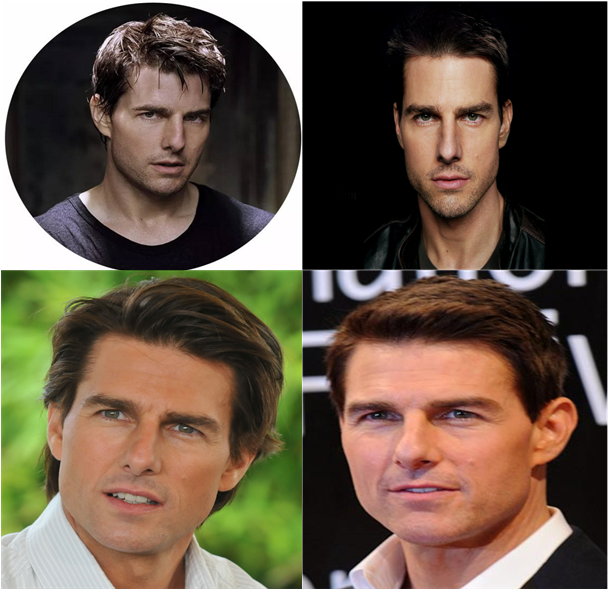}
  \caption{Examples of a type of facial images of Tom Cruise}
  \Description{Facial Images of Tom Cruise}
  \label{fig: figure8}
\end{figure}

Thus, we first constructed a binary classification sample set. 
Specifically, we designated the facial images of Tom Cruise as the target attributes, 
and those that were not his facial images as the non-target attributes. 
Figure~\ref{fig: figure8} shows some examples of a type of target attribute samples. 

Then, we trained the B-DNN model using the prepared sample set to realize the binary classification and achieve accurate identification of the target attributes. 

Meanwhile, in \textit{Discriminator} construction, concealment of malicious intent is realized by concealing the key. Therefore, in the B-DNN model in this paper, the key was dynamically concealed in a specific hidden layer. Moreover, this designated hidden was designed to have at least 128 neurons so that the output key could be no shorter than 128 bits to defend against reverse decryption. 
\begin{figure}[h]
  \centering
  \includegraphics[scale=0.25,trim=0mm 1mm 0mm 0mm,clip=true]{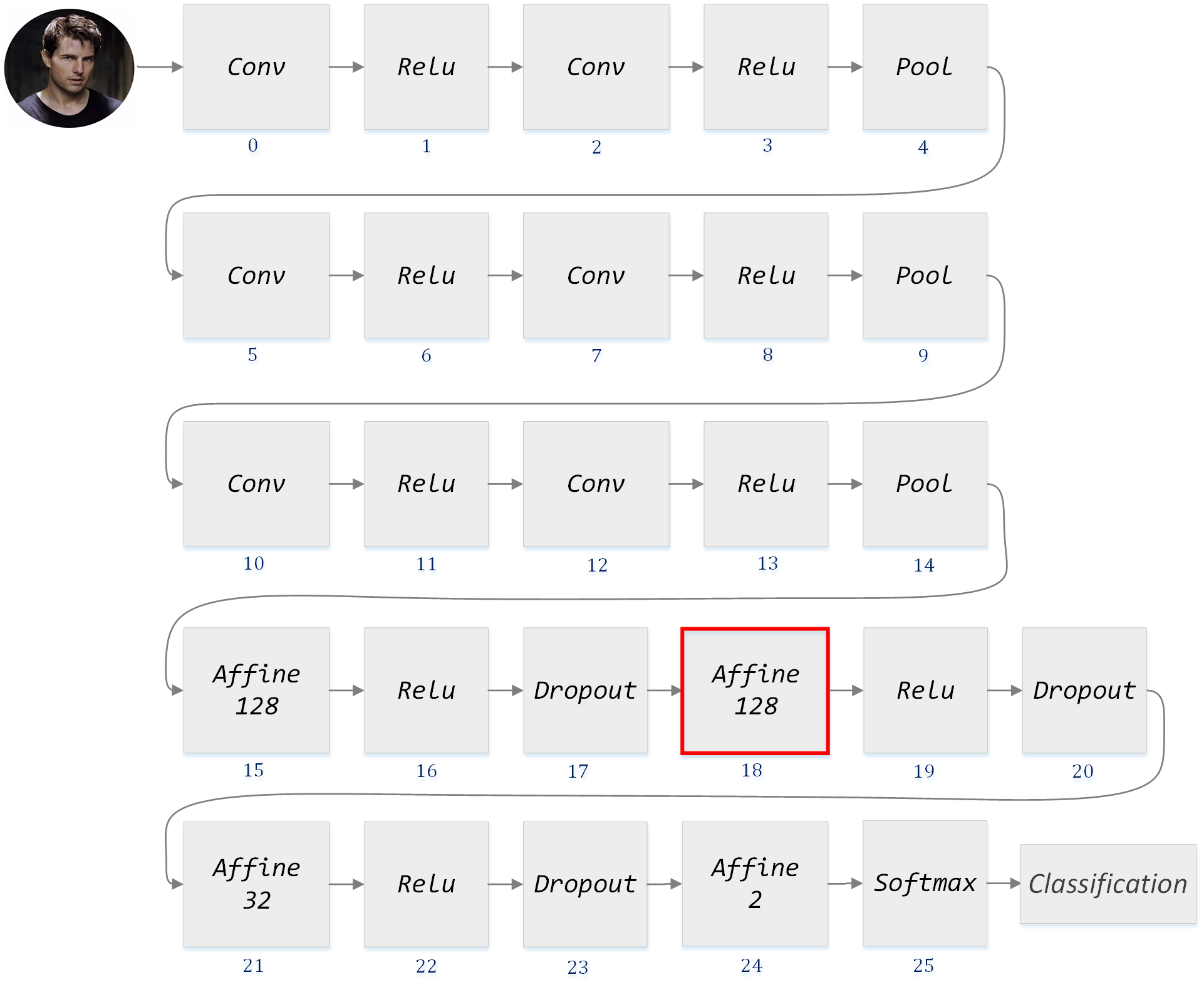}
  \caption{Unfolded layers of B-DNN}
  \Description{Layer Expansion of B-DNN}
  \label{fig: figure9}
\end{figure}

After repeated adjustment, training and testing, we obtained a B-DNN model (Figure~\ref{fig: figure9}). In the figure, \textit{Conv} refers to the convolutional layer, \textit{Relu} refers to the activation layer, which is implemented based on the rectified linear unit function; \textit{Pool} refers to the pooling layer, 
\textit{AffineX} refers to the fully-connected layer, with \textit{X} indicating the number of neurons within. 
For example, \textit{Affine128} refers to a fully-connected layer with 128 neurons. 
The \textit{Dropout} layer serves to prevent overfitting, and \textit{Softmax}, the last layer, performs normalization and classification.

\begin{figure}[h]
  \centering
  \includegraphics[scale=0.4,trim=0mm 1mm 0mm 0mm,clip=true]{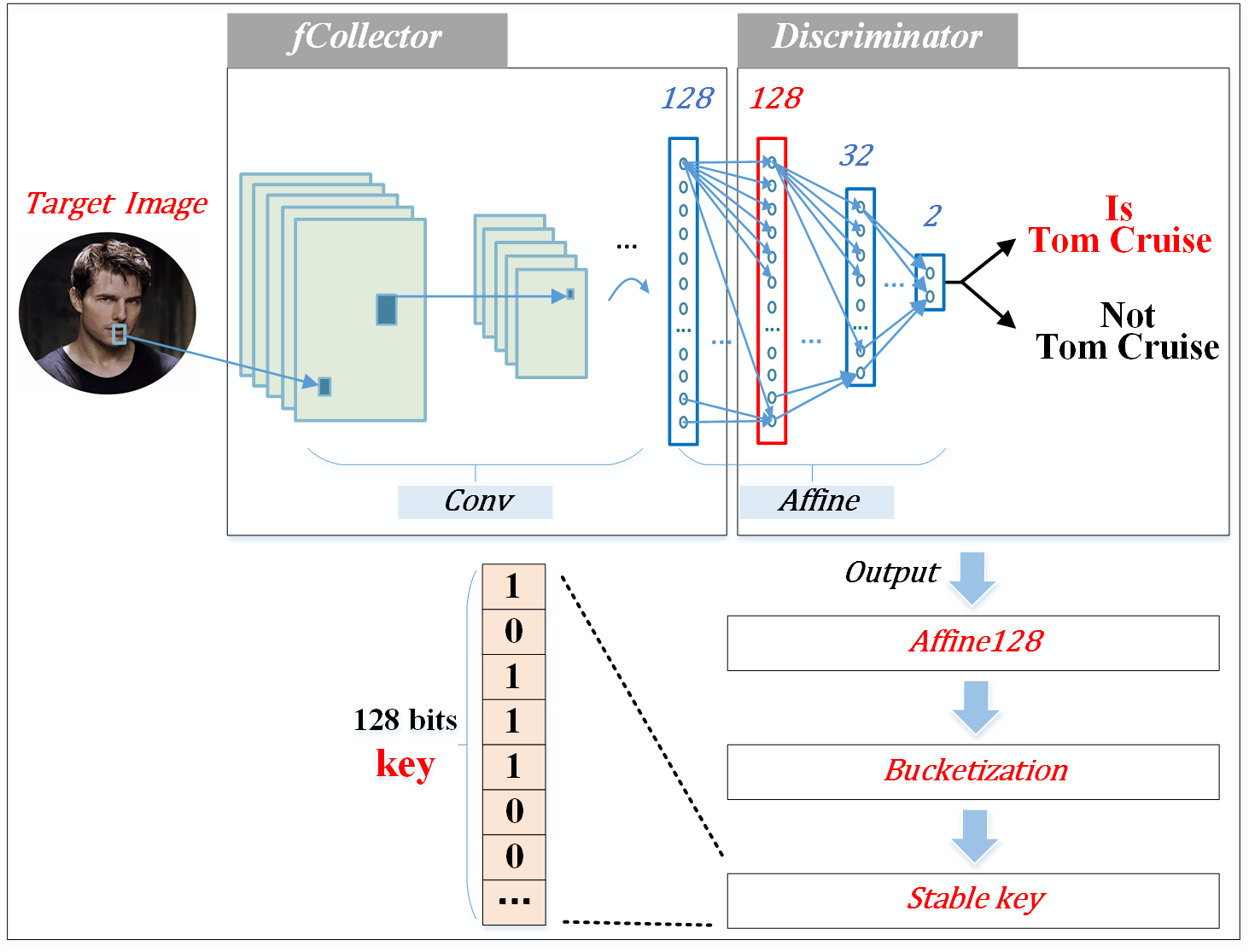}
  \caption{B-DNN Based Discriminator}
  \Description{B-DNN Based Discriminator}
  \label{fig: figure10}
\end{figure}

At last, we built a B-DNN based \textit{Discriminator} (Figure~\ref{fig: figure10}): Layers 0-17 in the model form the feature collector \textit{fCollector}, and Layers 18-25 constitute the \textit{Discriminator}. 
When the image of Tom Cruise is taken as the input, \textit{fCollector} will first collect the features, and the output of the fully-connected layer is taken as high dimension feature vector representation, 
which will then be input into the \textit{Discriminator} to identify whether the input image is the target attribute.

Specifically, the \textit{Discriminator} will produce output for \textit{Affine128} (Layer 18 shown in Figure~\ref{fig: figure9}), 
and the established B-DNN has stable representation on this output, i.e., it corresponds to a type of target attribute samples, and the 128-bit outputs on this layer are very similar. 
Then, we use the bucketization mechanism to process these similar outputs to achieve a stable output. 
For example, both 0.995 and 0.998 were output as 1, 0.0006 and 0.0001 were output as 0. 

Thereby, for a type of target attribute samples, the B-DNN based \textit{Discriminator} could produce a stable output, i.e., the stable key. 
Take for example the facial image of Tom Cruise shown in Figure~\ref{fig: figure10}; we finally obtained the stable key: 0x5c3871870e3c50f469dd86aeed38f7ed.

\subsection{Perceptual hash algorithm}
The perceptual hash algorithm generates a ``fingerprint'' character string for each image and then compares the fingerprints: 
a higher similarity between the fingerprints indicates more resemblance between the corresponding images. 
One notable feature of this algorithm is that it generates the fingerprint for a type of similar images. 
When an attacker designates a specific image, the perceptual hash algorithm can generate different embeddings for this image and keep their perceptual hash values the same. 
This considerably facilitates  coordinated adversarial attacks and helps circumvent detection of signatures and hashes by the defenders so that other bots will remain hidden when one bot is detected. 

\begin{figure}[!t]
  \centering
  \subfigure[$I_{0}$] {
    \includegraphics[scale=0.13,trim=0mm 0mm 0mm 0mm]{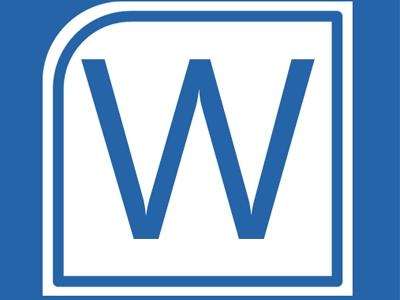}
  }
  \subfigure[not $I_{0}$] {
    \includegraphics[scale=0.344,trim=0mm 0mm 0mm 0mm]{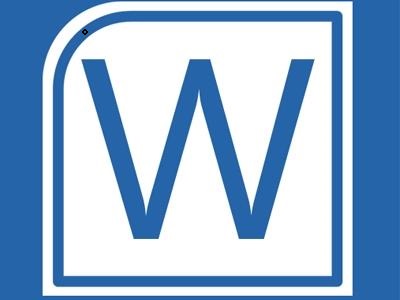}
  }
  \caption{Target images for Perceptual hash based Discriminator}
  \Description{Images of the same type generated based on the perceptual hash algorithm}
  \label{fig: figure11}
\end{figure}

Suppose we denote a type of images with perceptual hashes as \{$I_0,I_1,I_2,\dots,I_n$\}, 
and when one target image $I_0$ is designated, other images of the same type $\{I_1,I_2,\dots,I_n\}$ are usually man-made. Figure~\ref{fig: figure11}
 presents an example: Sub-fig. (a) shows a target image designated by the text, i.e., the image of the  Microsoft Word program. 
 We slightly altered (a) to generate Sub-fig. (b) while maintaining their visual resemblance. 
 They had distinct sha256 hash values but the same perceptual hash value. 
 Therefore, both sub-figs. (a) and (b) can be taken as the target image and be accurately identified by the perceptual hash-based \textit{Discriminator}.

\begin{figure}[h]
  \centering
  \includegraphics[scale=0.36,trim=0mm 1mm 0mm 0mm,clip=true]{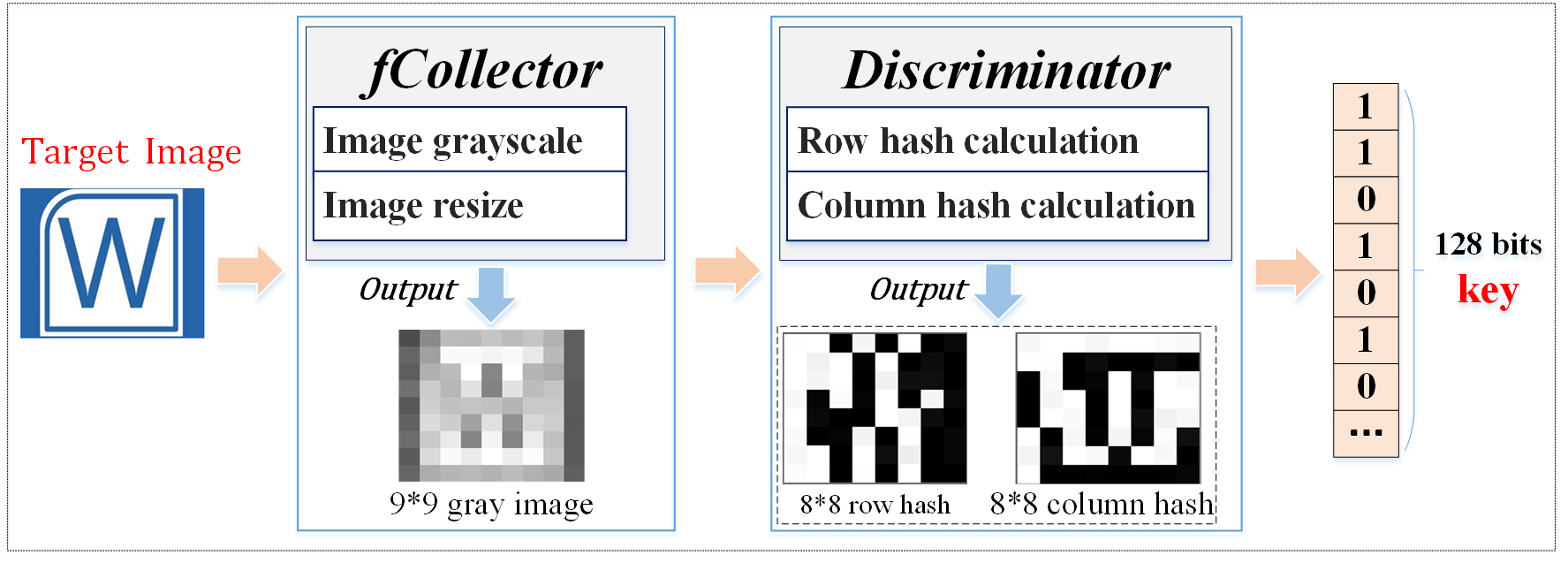}
  \caption{Perceptual hash algorithm based Discriminator}
  \Description{Perceptual hash algorithm based Discriminator}
  \label{fig: figure12}
\end{figure}

 In this paper, we constructed a \textit{Discriminator} based on the perceptual hash algorithm (Figure~\ref{fig: figure12}). 
 As with the B-DNN based \textit{Discriminator}, this \textit{Discriminator} is also aimed at the target attribute of unique-typedness and comprises \textit{fCollector} and \textit{Discriminator}. The \textit{fCollector}, with the two functions of ``image graying'' and ``image size adjustment'', collects the main features of an input image and outputs a 9*9 gray image as the feature image. 
 Then, the \textit{Discriminator} calculates the row hash and column hash of the feature image before outputting a hash image and a column hash image, both of a size of two 8*8. 
 These two output images correspond to two 64-bit hash calculation results, 
 which are then combined into a 128-bit output. 
 An example is shown in Figure~\ref{fig: figure12}: the target image $I_0$ in Figure~\ref{fig: figure11} is input into the \textit{Discriminator}, and the output is a 128-bit key, which is represented in hexadecimal as: 0xd4e8e8aa8c94d4d4ffc04b6b6baab680. 

To judge whether these four \textit{Discriminators} constructed above meet the requirements for UBM construction, we designed qualitative constraints on the ``unbreakable'' security attribute $Unb$ of the UBM according to Table~\ref{tab:table2}. 
That is, we specified the four critical values \textit{x}, \textit{y}, \textit{z} and \textit{w} in the tetrad \textit{eva}. 
Then, we specified the evaluation standards for non-enumerability of $P_{in}$ and $P_{out}$ as the constraint, and the evaluation standards for definiteness of $P_{sta}$ and $P_{acc}$ as the constraints. 
As per the formalized definition of UBM, $Unb$ is a two-tuple consisting of the two core features. 
That is to say, only when the malware constructed by the \textit{Discriminator} meets the evaluation standards of these two core features can it be defined as UBM, 
i.e., malware with the ``unbreakable'' security attribute $Unb$. 

To sum up, in order to construct the UBM, we would first evaluate the two core features of ``non-enumerability'' and ``definiteness'' in the upcoming text. 

\section{Evaluation}
This paper evaluated the \textit{Discriminators} by the following four aspects: (a) non-enumerability, (b) definiteness, (c) UBM volume increment, and (d) actual adversarial defense capability. Evaluation of the first two is a security evaluation of the \textit{Discriminator} to judge whether the discriminator meets the criteria for UBM construction; evaluation of the latter two is a fact evaluation of the applicability of the UBM. In other words, by evaluating (c) and (d), we could measure the UBM's capacity to survive antivirus engines and sound alarms for security defense.

\subsection{Evaluation of non-enumerability}
As per the standard for generating 128-bit unbreakable key, we set the limit of the infinite traversal at $2^{128}$. 
Thus, as shown in Table~\ref{tab:table2}, the enumeration space in both the input and output spaces in this paper had to exceed $2^{128}$. 
From the perspective of reverse definition, both of the two non-enumerability indicators $P_{in}$ and $P_{out}$ were set as $\leq 1/{2^{128}}$ in this paper. 

In this paper, all the four \textit{Discriminators} output key no shorter than 128 bits, and the key length was equivalent to the key strength because we used the AES algorithm. 
The \textit{Discriminator} constructed based on the typical hash algorithm output 256-bit key, and according to Formula~(\ref{con: equation4}), $P_{out}=1/{2^{256}}<1/{2^{128}}$; 
the rest three \textit{Discriminators} generated 128-bit key, and the corresponding 
$P_{out}=1/{2^{128}}$. 
That is to say, all of the four \textit{Discriminators} met the standards for unbreakable non-enumerability in the output space. 

\begin{enumerate}[$\bullet$]
  \item \textbf{Input space}
\end{enumerate}

As indicated by Formula~(\ref{con: equation2}) and ~(\ref{con: equation3}), to realize the unbreakable non-enumerability in the input space, the total number of input samples should be at least $2^{128}$ for the target attribute of uniqueness; 
and for the target attribute of unique-typedness, the total number of input sample types should be no less than $2^{128}$. 

For value transfer based \textit{Discriminator} and typical hash based \textit{Discriminator}, 
the input target attribute samples are text no shorter than 128 bits. There are no constraints on the content of the textual inputs, 
so the difficulty of traversing these inputs in the binary space will surely satisfy 
$P_{in}=1/{2^{len(target\ sample)}}\leq 1/{2^{128}}$
, 
where $len(target sample)$ refers to the binary length of the target attribute sample. 
It should be noted that the file type's target attributes also fall into the category of text. 

For B-DNN based \textit{Discriminator} and perceptual hash based \textit{Discriminator}, the input attribute samples were a type of images. When used to constructed the UBM, these two algorithms established a ``multiple-to-one'' mapping relation between a specific type of images and key no shorter than 128 bits, 
and generated the different outputs for different types of images. Since the enumeration space of the key were at least $2^{128}$ large, the number of corresponding types of images surely exceeded $2^{128}$, which met the standard
. 

To sum up, the input and output spaces of all the four \textit{Discriminators} constructed in this paper met the requirement of ``unbreakable'' non-enumerability. 

\subsection{Evaluation of definiteness}
The \textit{Discriminators} constructed based on the value transfer algorithm and typical hash algorithm establishes a ``one-to-one'' mapping relation between the target attribute and the key, and there is only one positive sample that corresponded to the target attribute. Therefore, according to Table~\ref{tab:table2}, $P_{sta}$  and $P_{acc}$ must be 1 and 0. 

Using Formula~(\ref{con: equation5}), we obtained that $P_{sta}$ was 1:1, which indicates a 100\% probability for ``key stability''. As shown in the non-enumerability evaluation, the input space comprised of positive and negative samples was infinitely large, which means that the number of negative samples corresponding to non-target attributes was no less than $2^{128}$. 
Therefore, according to Formula~(\ref{con: equation6}) and the ``one-to-one'' definite mapping relation, $P_{acc} \leq 0/{2^{128}}=0$, i.e., the ``key accessibility'' probability was 0. 

The \textit{Discriminators} constructed based on B-DNN and the perceptual hash algorithms establish a ``multiple-to-one'' mapping between the target attributes and the key. As shown by the eva tuple in Table~\ref{tab:table2}, we set an acceptable range of fault tolerance, requiring that $P_{sta}\geq 95\%$ and $P_{acc}\leq 0.5\%$.

For the B-DNN based \textit{Discriminator}, the input space is finite. According to Formula~(\ref{con: equation5}) and ~(\ref{con: equation6}), if the requirements for key stability and accessibility requirements can be satisfied in a finite input space, then they can also be satisfied in an actual input space that is infinitely large. 
Therefore, for the B-DNN based \textit{Discriminator}, the finite input space was created by the Tom Cruise's facial images data set (contains 1317 positive samples) and non-Tom Cruise's facial images data set (contains 1757 negative samples). Finally, this \textit{Discriminator} achieved a $P_{sta}$ of 98.5\%, and a $P_{acc}$ of 0, which met the evaluation standard for ``definiteness''.

Particularly, for the \textit{Discriminator} created based on the perceptual hash algorithm, all images in a same type of samples, except the initially designated image $I_0$, are inputs controlled and maliciously tweaked by the attacker, and they are invisible to the defender. The only visible image is the image $I_0$ in the input space. In this logic, though the discriminator built based on the perceptual hash algorithm is for the target attribute of unique-typedness, the ``multiple-to-one'' mapping relation it establishes is in essence equivalent to a ``one-to-one'' mapping relation. Then, by using Formula~(\ref{con: equation5}) and ~(\ref{con: equation6}), the $P_{sta}$ was 1 and the $P_{acc}$ was 0. 

To sum up, the four \textit{Discriminators} built in this paper all satisfied the requirements of ``key stability'' and ``key accessibility'', so they showed ``definiteness''. The results of non-enumerability evaluation also revealed that they met the standards. In other words, all the four \textit{Discriminators} fulfilled the ``unbreakable'' security attributes and hence met the standards for constructing the UBM. Then, based on the tuples presented in Table~\ref{tab:table2}, we constructed four UBMs in this paper: \textcircled{1} value transfer-based UBM, \textcircled{2} typical hash-based UBM, \textcircled{3} B-DNN based UBM, and \textcircled{4} perceptual hash-based UBM.

\subsection{UBM volume increment}
To minimize the chance of being suspected during malware attacks, the UBM volume increment must be considered. The UBM volume increment is defined as the combined volume of the rest four tuples among the six UBM tuples except for \textit{BenCode} and \textit{Unb}. 

First, the benign function code \textit{BenCode} implemented in this paper was a simple ``Hello, world'' pop-out. We packed the benign function codes into an executable application, whose volume was 20723 KB, i.e., circa 20.24 MB. 

Table~\ref{tab:table4} shows the volume increment of the four UBMs obtained based on the volume of \textit{BenCode}. 
Specifically, these UBMs were divided into two categories: UBMs constructed based on methods other than neural networks ("NN" for short), and those constructed based on NN. 
The former is a ``code-is-implementation'' method. Their UBM volume increment comes from the increase of codes due to the construction of \textit{Discriminator}, \textit{Judger} and \textit{Decryptor}, and the dynamic link library that these codes rely on during the program packing process.
We refer to this kind of UBM volume increment as ``code volume increment''. 
Construction of the latter category of UBM relies on training the neural network, adjusting model parameters, and saving of the architecture, which will bring in additional volume increment. 
In this paper, we refer to this kind of increment as ``non-code volume increment''.
\begin{table*}[]
  \caption{Volume increment of UBMs}
  \label{tab:table4}
  \begin{tabular}{@{}cccccc@{}}
  \toprule
  \multicolumn{2}{c}{\textbf{*** based UBM}}                             & \begin{tabular}[c]{@{}c@{}}\textbf{Code}\\\textbf{volume increment}\end{tabular} & \begin{tabular}[c]{@{}c@{}}\textbf{Non-code}\\\textbf{volume increment}\end{tabular} & \begin{tabular}[c]{@{}c@{}}\textbf{Total}\\\textbf{volume increment}\end{tabular} & \begin{tabular}[c]{@{}c@{}}\textbf{Volume of}\\\textbf{cipher payload}\end{tabular} \\ \midrule
  \multirow{3}{*}{\begin{tabular}[c]{@{}p{1.5cm}@{}}Non-NN method\end{tabular}} & Value transfer algorithm  & 21740KB-20723KB                & 0                                  & 0.99MB                          & \multirow{7}{*}{452KB}            \\ \cmidrule(lr){2-5}
                                             & Typical hash algorithm    & 21738KB-20723KB                & 0                                  & 0.99MB                          &                                   \\ \cmidrule(lr){2-5}
                                             & Perceptual hash algorithm      & 23245KB-20723KB                & 0                                  & 2.46MB                          &                                   \\ \cmidrule(lr){1-5}
  \begin{tabular}[c]{@{}p{1.5cm}@{}}NN method\end{tabular}   & B-DNN  & 23247KB-20723KB       & model: 1850KB                      & 4.27MB                          &                                   \\ \bottomrule
  \end{tabular}
\end{table*}

In Table~\ref{tab:table4}, the ``code volume increment'' and ``non-code volume increment'' columns sum up the total volume increment brought about by \textit{Discriminator}, \textit{Judger} and \textit{Decryptor}. 
There is another type of increment: the free volume increment brought about by the cipher payload \textit{cp}. 
As the type, function, or representation of the cipher payload varies, the volume of the free volume increment differs. 
In this connection, we analyzed the volumes of payloads labeled ``malicious'' in the public cloud sandbox  
and concluded that the volume of almost half of payloads is smaller than 1 MB. As stated in Table~\ref{tab:table2}, 
when constructing the UBM, we took the a.exe in trickbot as the malicious payload, whose volume stayed at 452 KB before and after encryption. 
Compared with the volume of a benign function program, the volume of the payload is negligible.

Finally, based on the data recorded in Table~\ref{tab:table4}, we reached the following conclusion: 
The non-NN method brought about little volume increment when used to construct the malware. Thus, it would make the malware less likely to be suspected by the defender in actual applications. 
As for the NN method, the network model that used to construct the B-DNN based UBM in this paper was artificially built without relying on the open source TensorFlow library, so the model can be saved to a very small size, making the final total volume increment stay at only 4.27 MB. 
Compared with that of a benign function program, this volume is still unlikely to raise the defender's suspicion.

Given the analyses of volume increment above, the range of volume increment of the four UBMs built in this paper was [0.99 MB, 4.27 MB], and the larger the volume of the benign function program was, the less likely the UBM would be detected as suspicious. 
Therefore, in actual attack scenarios, selecting proper benign function programs is necessary according to the acceptable range of malware volume increments when constructing the UBM. 

\subsection{Actual anti-defense capability}
From the perspective of adversarial attack and defense, the attacker conceals the malware to survive or circumvent the detection of the defender's antivirus engine, thereby creating a chance to launch attacks. 
To better demonstrate the actual attack performance of the UBM, we used antivirus engines deployed on the cloud and on the host computer to test the anti-defense capacity of our constructed UBMs. 
One antivirus engine on the host computer and nearly 100 antivirus engines on the cloud were employed to conduct detection.
Especially VirusTotal \cite{vt13}, which integrates most of the well-known anti-virus engines, can perform static analysis or dynamic detection. Therefore, the detection result delivered by VirusTotal is convincing, and the same is true for ThreatBook \cite{threatbook14}, as shown in Table~\ref{tab:table5}.

As mentioned before, the plain payload used in this paper is a.exe in trickbot. 
When it was downloaded to the host computer or uploaded to the cloud detection engine, the antivirus engine on the host computer would immediately scan and clean it. 
The VirusTotal employed on the cloud side conducted detection using 72 antivirus engines, 57 of which detected plain payloads as ``malicious''. Among the 25 antivirus engines in ThreatBook, 
nine detected plain payloads as ``malicious'' and marked their threat level as ``malicious''. 

Unlike the plain payload, the cipher payload was identified as ``secure'' files by antivirus engines both on the host computer and the cloud. 
Moreover, as shown in Table~\ref{tab:table5}, the four complete and releasable UBMs constructed in this paper have also been detected. 
The result shows that almost all four UBMs were identified as ``safe''. 
Therefore, we could conclude that even when the UBM or the cipher payload was open source, the defender still could not find a way to crack the malicious intent hidden in the malware by forward or reverse analysis methods. 
Finally, it should be noted that in Table~\ref{tab:table5}, as the file size upper limit was set at 20 MB on ThreatBook, while our uploaded UBM exceeded this limit, so we did not use the engines on ThreatBook to test our constructed UBMs. 
\begin{table}[!t]
  \caption{Detection results of antivirus engines}
  \label{tab:table5}
  \begin{tabular}{@{}ccccc@{}}
  \toprule
  \multicolumn{2}{c}{\begin{tabular}[c]{@{}c@{}} \textbf{Antivirus} \\ \textbf{engine}\end{tabular}} & \begin{tabular}[c]{@{}c@{}} \textbf{plain}\\ \textbf{payload}\end{tabular} & \begin{tabular}[c]{@{}c@{}}\textbf{cipher}\\ \textbf{payload}\end{tabular} & \begin{tabular}[c]{@{}c@{}}\textbf{UBM}\\ \end{tabular} \\ \midrule
  Host                                 & huorong                                  & Kill immediately                                        & 0 risk                                                   & 0 risk                                                              \\ \midrule
  \multirow{5}{*}{Cloud}               & \multirow{4}{*}{VirusTotal}              & \multirow{4}{*}{57/72}                                  & \multirow{4}{*}{0/61}                                  & \textcircled{\scriptsize 1}: 1/72                             \\ \cmidrule(l){5-5}
                                       &                                          &                                                         &                                                          & \textcircled{\scriptsize 2}: 1/71                            \\ \cmidrule(l){5-5} 
                                       &                                          &                                                         &                                                          & \textcircled{\scriptsize 3}: 1/70                            \\ \cmidrule(l){5-5}
                                       &                                          &                                                         &                                                          & \textcircled{\scriptsize 4}: 1/71                            \\ \cmidrule(l){2-5}
                                       & ThreatBook                               & 9/25                                                    & 0/25                                                   & --                                                                    \\ \midrule
  \multicolumn{5}{l}{\tiny \begin{tabular}[c]{@{}p{7.8cm}@{}}\textbf{Note$\colon$} \textcircled{\scriptsize 1}Value transfer algorithm based UBM, \textcircled{\scriptsize 2}Typical hash algorithm based UBM, \textcircled{\scriptsize 3}B-DNN based UBM, \textcircled{\scriptsize 4}Perceptual hash algorithm based UBM.\end{tabular}}    \\ \bottomrule
  \end{tabular}
\end{table}

To sum up, by evaluating the four aspects of the constructed UBMs, i.e., the non-enumerability, definiteness, UBM volume increment and actual anti-defense capability, we confirmed the security benefits of UBM. 
To be specific, by evaluating the ``non-enumerability'' and ``key stability and key accessibility'' of the UBMs, we verified their ``unbreakable'' security attribute; evaluation of their ``UBM volume increment'' and ``actual anti-defense capability'' attested that the UBMs could conceal the malicious intent and survive/circumvent detection by almost all available well-known antivirus engines for the time being. 

\section{Defense against Unbreakable Malware}
In this section, we probed into possible defense measures based on analysis of vulnerabilities of UBM revealed during target attribute designation and in the functioning stages, with a vision to address new security threats incurred by UBM.
\begin{enumerate}[$\bullet$]
  \item \textbf{Vulnerability incurred by target attribute designation}
\end{enumerate}

In the input space of the \textit{Discriminator}, there were various types of target attributes that can be designated and connected to the victim host computer, such as the software environment, physical environment, user behavior and geographic location, as shown in Figure~\ref{fig: figure4}. By selecting proper target attributes, the defender can have two possible defense solutions. 

First, from the perspective of security defense, the intrinsic attributes of the victim, such as computer name and user name, can be utilized. In the input space, these intrinsic attributes help to ensure non-enumerability; on the defender's side, however, calling these attributes can be taken as features that helps to detect UBM.  

As for the second solution, when the designated target attribute is a specific file or image, this type of target attributes will see a considerable traversal space in the victim host computer (this space is a subset of the input space). In this case, the \textit{Discriminator} needs to frequently execute ``read'' operations (loading pictures frequently, for instance) in an attempt to decrypt the malicious payload. On the defender's side, these frequent operations are aberrant behavior and can be taken as a feature for the detection of UBM.

\begin{enumerate}[$\bullet$]
  \item \textbf{Vulnerability incurred in the functioning stage}
\end{enumerate}

Based on the malware attack chain \cite{tian01}, we focused on malware concealment capability of our constructed UBMs before an attack was unleashed, while the two stages of ``effect'' and ``command and control'' were not considered. However, even if the UBM has achieved nearly complete concealment before launching an attack, it will still be subject to detection by process behavior-based \cite{DBLP:conf/compsac/TobiyamaYSIY16,DBLP:journals/compsec/RhodeBJ18} and malicious traffic-based \cite{homayoun2018botshark,DBLP:conf/sp/MarinCC19} detections after launching the attack. Therefore, the defender can defend against the UBM by real-time scanning and removal of malicious behaviors.

\section{Conclusion and Future work}
For the first time, this paper proposed the concept of unbreakable malware -- UBM, and made a systematic probe into it through modeling, method analysis, experiments, evaluation and anti-defense tests. Compared with existing methods, we built UBM and thus summarized the features of this type of malware: any malware that satisfies the two core features of ``non-enumerability'' and ``definiteness'' possesses the ``unbreakable'' security attribute.  

By formalized modeling, research on security attributes, architecture building, and exploration of evaluation standards, we constructed UBM instances based on four algorithms. In the experiments, we verified the ``unbreakable'' security attribute of the constructed UBMs through security evaluation; meanwhile, by evaluating the applicability of the UBM, we confirmed that UBM is a novel high-threat attack technique. That is, it has a volume increment that is hardly perceivable, and can get away with the detection by nearly one hundred antivirus engines deployed locally or on the cloud. 

Moreover, we have explored possible defense solutions to attacks of UBM and uncovered some of its defects. In future work, we will try to repair these defects and improve the performance of UBM. More importantly, we will invest more to explore defense solutions against this type of advanced ``unbreakable'' malware.  

\bibliographystyle{ACM-Reference-Format}
\bibliography{sample-base-DLpaper}


\begin{thebibliography}{18}


\ifx \showCODEN    \undefined \def \showCODEN     #1{\unskip}     \fi
\ifx \showDOI      \undefined \def \showDOI       #1{#1}\fi
\ifx \showISBNx    \undefined \def \showISBNx     #1{\unskip}     \fi
\ifx \showISBNxiii \undefined \def \showISBNxiii  #1{\unskip}     \fi
\ifx \showISSN     \undefined \def \showISSN      #1{\unskip}     \fi
\ifx \showLCCN     \undefined \def \showLCCN      #1{\unskip}     \fi
\ifx \shownote     \undefined \def \shownote      #1{#1}          \fi
\ifx \showarticletitle \undefined \def \showarticletitle #1{#1}   \fi
\ifx \showURL      \undefined \def \showURL       {\relax}        \fi
\providecommand\bibfield[2]{#2}
\providecommand\bibinfo[2]{#2}
\providecommand\natexlab[1]{#1}
\providecommand\showeprint[2][]{arXiv:#2}

\bibitem[\protect\citeauthoryear{Blackthorne, Bulazel, Fasano, Biernat, and
  Yener}{Blackthorne et~al\mbox{.}}{2016}]%
        {DBLP:conf/woot/BlackthorneBFBY16}
\bibfield{author}{\bibinfo{person}{Jeremy Blackthorne}, \bibinfo{person}{Alexei
  Bulazel}, \bibinfo{person}{Andrew Fasano}, \bibinfo{person}{Patrick Biernat},
  {and} \bibinfo{person}{B{\"{u}}lent Yener}.} \bibinfo{year}{2016}\natexlab{}.
\newblock \showarticletitle{AVLeak: Fingerprinting Antivirus Emulators through
  Black-Box Testing}. In \bibinfo{booktitle}{\emph{10th {USENIX} Workshop on
  Offensive Technologies, {WOOT} 16, Austin, TX, USA, August 8-9, 2016}},
  \bibfield{editor}{\bibinfo{person}{Natalie Silvanovich} {and}
  \bibinfo{person}{Patrick Traynor}} (Eds.). \bibinfo{publisher}{{USENIX}
  Association}.
\newblock
\urldef\tempurl%
\url{https://www.usenix.org/conference/woot16/workshop-program/presentation/blackthorne}
\showURL{%
\tempurl}


\bibitem[\protect\citeauthoryear{Chen and Abu{-}Nimeh}{Chen and
  Abu{-}Nimeh}{2011}]%
        {DBLP:journals/computer/ChenA11}
\bibfield{author}{\bibinfo{person}{Thomas~M. Chen} {and} \bibinfo{person}{Saeed
  Abu{-}Nimeh}.} \bibinfo{year}{2011}\natexlab{}.
\newblock \showarticletitle{Lessons from Stuxnet}.
\newblock \bibinfo{journal}{\emph{Computer}} \bibinfo{volume}{44},
  \bibinfo{number}{4} (\bibinfo{year}{2011}), \bibinfo{pages}{91--93}.
\newblock
\urldef\tempurl%
\url{https://doi.org/10.1109/MC.2011.115}
\showDOI{\tempurl}


\bibitem[\protect\citeauthoryear{Clement}{Clement}{2019}]%
        {David03}
\bibfield{author}{\bibinfo{person}{David Clement}.}
  \bibinfo{year}{2019}\natexlab{}.
\newblock \bibinfo{booktitle}{\emph{2019 Midyear Security Roundup: Evasive
  Threats, Pervasive Effects}}.
\newblock \bibinfo{type}{{T}echnical {R}eport}.
\newblock
\urldef\tempurl%
\url{https://documents.trendmicro.com/assets/rpt/rpt-evasive-threats-pervasive-effects.pdf}
\showURL{%
\tempurl}


\bibitem[\protect\citeauthoryear{google}{google}{[n. d.]}]%
        {vt13}
google \bibinfo{year}{[n. d.]}\natexlab{}.
\newblock \bibinfo{title}{VirusTotal}.
\newblock
\newblock
\urldef\tempurl%
\url{https://www.virustotal.com/gui/home/upload}
\showURL{%
Retrieved August 18, 2020 from \tempurl}


\bibitem[\protect\citeauthoryear{Ji, Fang, Cui, Wang, Gan, Han, and Yu}{Ji
  et~al\mbox{.}}{2020}]%
        {tian01}
\bibfield{author}{\bibinfo{person}{Tiantian Ji}, \bibinfo{person}{Binxing
  Fang}, \bibinfo{person}{Xiang Cui}, \bibinfo{person}{Zhongru Wang},
  \bibinfo{person}{Ruiling Gan}, \bibinfo{person}{Yu Han}, {and}
  \bibinfo{person}{Weiqiang Yu}.} \bibinfo{year}{2020}\natexlab{}.
\newblock \showarticletitle{Research on deep learning-powered malware attack
  and defense techniques}.
\newblock \bibinfo{journal}{\emph{Chinese Journal of Computers}}
  \bibinfo{volume}{43}, \bibinfo{number}{13} (\bibinfo{date}{May}
  \bibinfo{year}{2020}), \bibinfo{pages}{1--29}.
\newblock
\urldef\tempurl%
\url{http://cjc.ict.ac.cn/online/bfpub/jtt-202058150207.pdf}
\showURL{%
\tempurl}


\bibitem[\protect\citeauthoryear{Kirat, Jang, and Stoecklin}{Kirat
  et~al\mbox{.}}{2018}]%
        {Dhilung04}
\bibfield{author}{\bibinfo{person}{Dhilung Kirat}, \bibinfo{person}{Jiyong
  Jang}, {and} \bibinfo{person}{Marc~Ph. Stoecklin}.}
  \bibinfo{year}{2018}\natexlab{}.
\newblock \showarticletitle{Deeplocker - concealing targeted attacks with ai
  locksmithing}. In \bibinfo{booktitle}{\emph{Black Hat USA, 2018, Las Vegas,
  USA, August 4-9, 2018}}. \bibinfo{publisher}{informatech},
  \bibinfo{pages}{3873--3878}.
\newblock
\urldef\tempurl%
\url{https://www.blackhat.com/us-18/briefings/schedule/index.html#deeplocker---concealing-targeted-attacks-with-ai-locksmithing-11549}
\showURL{%
\tempurl}


\bibitem[\protect\citeauthoryear{Kolbitsch, Kirda, and Kruegel}{Kolbitsch
  et~al\mbox{.}}{2011}]%
        {DBLP:conf/ccs/KolbitschKK11}
\bibfield{author}{\bibinfo{person}{Clemens Kolbitsch}, \bibinfo{person}{Engin
  Kirda}, {and} \bibinfo{person}{Christopher Kruegel}.}
  \bibinfo{year}{2011}\natexlab{}.
\newblock \showarticletitle{The power of procrastination: detection and
  mitigation of execution-stalling malicious code}. In
  \bibinfo{booktitle}{\emph{Proceedings of the 18th {ACM} Conference on
  Computer and Communications Security, {CCS} 2011, Chicago, Illinois, USA,
  October 17-21, 2011}}, \bibfield{editor}{\bibinfo{person}{Yan Chen},
  \bibinfo{person}{George Danezis}, {and} \bibinfo{person}{Vitaly Shmatikov}}
  (Eds.). \bibinfo{publisher}{{ACM}}, \bibinfo{pages}{285--296}.
\newblock
\urldef\tempurl%
\url{https://doi.org/10.1145/2046707.2046740}
\showDOI{\tempurl}


\bibitem[\protect\citeauthoryear{Mar{\'{\i}}n, Casas, and
  Capdehourat}{Mar{\'{\i}}n et~al\mbox{.}}{2019}]%
        {DBLP:conf/sp/MarinCC19}
\bibfield{author}{\bibinfo{person}{Gonzalo Mar{\'{\i}}n},
  \bibinfo{person}{Pedro Casas}, {and} \bibinfo{person}{Germ{\'{a}}n
  Capdehourat}.} \bibinfo{year}{2019}\natexlab{}.
\newblock \showarticletitle{Deep in the Dark - Deep Learning-Based Malware
  Traffic Detection Without Expert Knowledge}. In
  \bibinfo{booktitle}{\emph{2019 {IEEE} Security and Privacy Workshops, {SP}
  Workshops 2019, San Francisco, CA, USA, May 19-23, 2019}}.
  \bibinfo{publisher}{{IEEE}}, \bibinfo{pages}{36--42}.
\newblock
\urldef\tempurl%
\url{https://doi.org/10.1109/SPW.2019.00019}
\showDOI{\tempurl}


\bibitem[\protect\citeauthoryear{Miramirkhani, Appini, Nikiforakis, and
  Polychronakis}{Miramirkhani et~al\mbox{.}}{2017}]%
        {DBLP:conf/sp/MiramirkhaniANP17}
\bibfield{author}{\bibinfo{person}{Najmeh Miramirkhani},
  \bibinfo{person}{Mahathi~Priya Appini}, \bibinfo{person}{Nick Nikiforakis},
  {and} \bibinfo{person}{Michalis Polychronakis}.}
  \bibinfo{year}{2017}\natexlab{}.
\newblock \showarticletitle{Spotless Sandboxes: Evading Malware Analysis
  Systems Using Wear-and-Tear Artifacts}. In \bibinfo{booktitle}{\emph{2017
  {IEEE} Symposium on Security and Privacy, {SP} 2017, San Jose, CA, USA, May
  22-26, 2017}}. \bibinfo{publisher}{{IEEE} Computer Society},
  \bibinfo{pages}{1009--1024}.
\newblock
\urldef\tempurl%
\url{https://doi.org/10.1109/SP.2017.42}
\showDOI{\tempurl}


\bibitem[\protect\citeauthoryear{Rhode, Burnap, and Jones}{Rhode
  et~al\mbox{.}}{2018}]%
        {DBLP:journals/compsec/RhodeBJ18}
\bibfield{author}{\bibinfo{person}{Matilda Rhode}, \bibinfo{person}{Pete
  Burnap}, {and} \bibinfo{person}{Kevin Jones}.}
  \bibinfo{year}{2018}\natexlab{}.
\newblock \showarticletitle{Early-stage malware prediction using recurrent
  neural networks}.
\newblock \bibinfo{journal}{\emph{Comput. Secur.}}  \bibinfo{volume}{77}
  (\bibinfo{year}{2018}), \bibinfo{pages}{578--594}.
\newblock
\urldef\tempurl%
\url{https://doi.org/10.1016/j.cose.2018.05.010}
\showDOI{\tempurl}


\bibitem[\protect\citeauthoryear{Sajad, Marzieh, Sattar, Ali, and Raouf}{Sajad
  et~al\mbox{.}}{2018}]%
        {homayoun2018botshark}
\bibfield{author}{\bibinfo{person}{Homayoun Sajad}, \bibinfo{person}{Ahmadzadeh
  Marzieh}, \bibinfo{person}{Hashemi Sattar}, \bibinfo{person}{Dehghantanha
  Ali}, {and} \bibinfo{person}{Khayami Raouf}.}
  \bibinfo{year}{2018}\natexlab{}.
\newblock \showarticletitle{BoTShark: A deep learning approach for botnet
  traffic detection}.
\newblock In \bibinfo{booktitle}{\emph{Cyber Threat Intelligence}},
  \bibfield{editor}{\bibinfo{person}{Dehghantanha Ali}, \bibinfo{person}{Conti
  Mauro}, {and} \bibinfo{person}{Dargahi Tooska}} (Eds.).
  \bibinfo{series}{Advances in Information Security},
  Vol.~\bibinfo{volume}{70}. \bibinfo{publisher}{Springer, Cham},
  \bibinfo{pages}{137--153}.
\newblock
\urldef\tempurl%
\url{https://doi.org/10.1007/978-3-319-73951-9_7}
\showDOI{\tempurl}


\bibitem[\protect\citeauthoryear{Stephens, Grosen, Salls, Dutcher, Wang,
  Corbetta, Shoshitaishvili, Kruegel, and Vigna}{Stephens
  et~al\mbox{.}}{2016}]%
        {DBLP:conf/ndss/StephensGSDWCSK16}
\bibfield{author}{\bibinfo{person}{Nick Stephens}, \bibinfo{person}{John
  Grosen}, \bibinfo{person}{Christopher Salls}, \bibinfo{person}{Andrew
  Dutcher}, \bibinfo{person}{Ruoyu Wang}, \bibinfo{person}{Jacopo Corbetta},
  \bibinfo{person}{Yan Shoshitaishvili}, \bibinfo{person}{Christopher Kruegel},
  {and} \bibinfo{person}{Giovanni Vigna}.} \bibinfo{year}{2016}\natexlab{}.
\newblock \showarticletitle{Driller: Augmenting Fuzzing Through Selective
  Symbolic Execution}. In \bibinfo{booktitle}{\emph{23rd Annual Network and
  Distributed System Security Symposium, {NDSS} 2016, San Diego, California,
  USA, February 21-24, 2016}}. \bibinfo{publisher}{The Internet Society}.
\newblock
\urldef\tempurl%
\url{http://wp.internetsociety.org/ndss/wp-content/uploads/sites/25/2017/09/driller-augmenting-fuzzing-through-selective-symbolic-execution.pdf}
\showURL{%
\tempurl}


\bibitem[\protect\citeauthoryear{The MITRE Corporation}{The MITRE
  Corporation}{2020}]%
        {mitre12}
The MITRE Corporation \bibinfo{year}{2015--2020}\natexlab{}.
\newblock \bibinfo{title}{MITRE ATT\&CK}.
\newblock
\newblock
\urldef\tempurl%
\url{https://attack.mitre.org/}
\showURL{%
Retrieved August 18, 2020 from \tempurl}


\bibitem[\protect\citeauthoryear{ThreatBook}{ThreatBook}{[n. d.]}]%
        {threatbook14}
ThreatBook \bibinfo{year}{[n. d.]}\natexlab{}.
\newblock \bibinfo{title}{Threatbook cloud sandbox}.
\newblock
\newblock
\urldef\tempurl%
\url{https://s.threatbook.cn/}
\showURL{%
Retrieved August 18, 2020 from \tempurl}


\bibitem[\protect\citeauthoryear{Tobiyama, Yamaguchi, Shimada, Ikuse, and
  Yagi}{Tobiyama et~al\mbox{.}}{2016}]%
        {DBLP:conf/compsac/TobiyamaYSIY16}
\bibfield{author}{\bibinfo{person}{Shun Tobiyama}, \bibinfo{person}{Yukiko
  Yamaguchi}, \bibinfo{person}{Hajime Shimada}, \bibinfo{person}{Tomonori
  Ikuse}, {and} \bibinfo{person}{Takeshi Yagi}.}
  \bibinfo{year}{2016}\natexlab{}.
\newblock \showarticletitle{Malware Detection with Deep Neural Network Using
  Process Behavior}. In \bibinfo{booktitle}{\emph{40th {IEEE} Annual Computer
  Software and Applications Conference, {COMPSAC} Workshops 2016, Atlanta, GA,
  USA, June 10-14, 2016}}. \bibinfo{publisher}{{IEEE} Computer Society},
  \bibinfo{pages}{577--582}.
\newblock
\urldef\tempurl%
\url{https://doi.org/10.1109/COMPSAC.2016.151}
\showDOI{\tempurl}


\bibitem[\protect\citeauthoryear{Ubale and Jain}{Ubale and Jain}{2020}]%
        {DBLP:books/sp/20/UbaleJ20}
\bibfield{author}{\bibinfo{person}{Tushar Ubale} {and}
  \bibinfo{person}{Ankit~Kumar Jain}.} \bibinfo{year}{2020}\natexlab{}.
\newblock \showarticletitle{Survey on DDoS Attack Techniques and Solutions in
  Software-Defined Network}.
\newblock In \bibinfo{booktitle}{\emph{Handbook of Computer Networks and Cyber
  Security, Principles and Paradigms}},
  \bibfield{editor}{\bibinfo{person}{Brij~B. Gupta},
  \bibinfo{person}{Gregorio~Mart{\'{\i}}nez P{\'{e}}rez},
  \bibinfo{person}{Dharma~P. Agrawal}, {and} \bibinfo{person}{Deepak Gupta}}
  (Eds.). \bibinfo{publisher}{Springer}, \bibinfo{pages}{389--419}.
\newblock
\urldef\tempurl%
\url{https://doi.org/10.1007/978-3-030-22277-2\_15}
\showDOI{\tempurl}


\bibitem[\protect\citeauthoryear{Wang, Hassan, Li, Jee, Yu, Zou, Rhee, Chen,
  Cheng, Gunter, and Chen}{Wang et~al\mbox{.}}{2020}]%
        {DBLP:conf/ndss/WangHLJYZRCCGC20}
\bibfield{author}{\bibinfo{person}{Qi Wang}, \bibinfo{person}{Wajih~Ul Hassan},
  \bibinfo{person}{Ding Li}, \bibinfo{person}{Kangkook Jee},
  \bibinfo{person}{Xiao Yu}, \bibinfo{person}{Kexuan Zou},
  \bibinfo{person}{Junghwan Rhee}, \bibinfo{person}{Zhengzhang Chen},
  \bibinfo{person}{Wei Cheng}, \bibinfo{person}{Carl~A. Gunter}, {and}
  \bibinfo{person}{Haifeng Chen}.} \bibinfo{year}{2020}\natexlab{}.
\newblock \showarticletitle{You Are What You Do: Hunting Stealthy Malware via
  Data Provenance Analysis}. In \bibinfo{booktitle}{\emph{27th Annual Network
  and Distributed System Security Symposium, {NDSS} 2020, San Diego,
  California, USA, February 23-26, 2020}}. \bibinfo{publisher}{The Internet
  Society}.
\newblock
\urldef\tempurl%
\url{https://www.ndss-symposium.org/ndss-paper/you-are-what-you-do-hunting-stealthy-malware-via-data-provenance-analysis/}
\showURL{%
\tempurl}


\bibitem[\protect\citeauthoryear{Zalewski}{Zalewski}{2020}]%
        {afl11}
\bibfield{author}{\bibinfo{person}{Micha? Zalewski}.}
  \bibinfo{year}{2015--2020}\natexlab{}.
\newblock \bibinfo{title}{American fuzzy lop}.
\newblock
\newblock
\urldef\tempurl%
\url{https://lcamtuf.coredump.cx/afl/}
\showURL{%
Retrieved August 18, 2020 from \tempurl}


\end{thebibliography}

\end{document}